\documentclass{svjour3}
\pdfoutput=1
\usepackage{etex}
\usepackage{comment}
\journalname{Special Issue on Automation of Software Test}

\author{Vincenzo Musco   \and
        Martin Monperrus \and
        Philippe Preux}

\institute{Vincenzo Musco \at
              University of Lille, CRIStAL and INRIA \\
              \email{vincenzo.musco@inria.fr}
           \and
           Martin Monperrus \at
           University of Lille, CRIStAL and INRIA \\
           \email{martin.monperrus@univ-lille1.fr}
           \and
           Philippe Preux \at
           University of Lille, CRIStAL and INRIA \\
           \email{philippe.preux@univ-lille.fr}
}

\date{First Online: 26 July 2016}

\smartqed

\newcommand{\figureinlatex}{true}

\RequirePackage{fix-cm}

\usepackage[utf8]{inputenc}
\usepackage{amssymb}
\usepackage{hyperref} 
\usepackage{url}
\usepackage{color}
\usepackage{booktabs}
\usepackage{paralist}
\usepackage{xspace}
\usepackage{ifthen}
\usepackage{listings}
\usepackage{afterpage}
\usepackage{graphicx}
\usepackage{pdflscape}
\usepackage{comment}
\usepackage{multirow}
\usepackage{tabularx}
\usepackage{rotating}
\usepackage{mathrsfs}
\usepackage{framed}
\usepackage{xcolor}
\usepackage{geometry}
\usepackage[boxed,ruled,vlined,linesnumbered]{algorithm2e}
\usepackage{graphicx}

\newcounter{rqs}
\newcommand{\RQ}[1]{\refstepcounter{rqs}\vspace*{3mm}\textit{\underline{Research Question \arabic{rqs}} #1}\vspace*{3mm}}
\newcommand{\RQInline}[1]{\refstepcounter{rqs}\vspace*{3mm}\textit{\underline{Research Question \arabic{rqs}} #1}}
\newcommand{\RQRESET}{\setcounter{rqs}{0}}

\renewcommand{\S}{{\cal S}}
\renewcommand{\O}{{\cal O}}
\newcommand{\U}{{\cal U}}
\newcommand{\D}{{\cal D}}
\newcommand{\C}{{\cal C}}
\newcommand{\N}{{\cal N}}
\newcommand{\K}{{\cal K}}

\newcommand{\eg}{\textit{e.g. }}
\newcommand{\ie}{\textit{i.e. }}
\newcommand{\etc}{\textit{etc.}}
\newcommand{\cf}{\textit{cf. }}
\newcommand{\aka}{\textit{a.k.a. }}
\newcommand{\etal}{\textit{et al.}\xspace}
\newcommand{\resp}{\textit{resp.}\xspace}

\newcommand{\cgalone}{\mathscr{C}}
\newcommand{\cg}{$\cgalone_{B}$\xspace}
\newcommand{\cgj}{$\cgalone_{S}$\xspace}
\newcommand{\cgcha}{$\cgalone_{H}$\xspace}
\newcommand{\cgf}{$\cgalone_{F}$\xspace}

\newcommand{\takeaway}[1]{\begin{framed}#1\end{framed}}

\newcommand{\nbprojects}{10\xspace}
\newcommand{\nbmutants}{17,000\xspace}

\begin{document}

\title{A Large-Scale Study of Call Graph-based Impact Prediction using Mutation Testing}

\maketitle

\begin{abstract}
In software engineering, impact analysis involves predicting the software elements (\eg modules, classes, methods) potentially impacted by a change in the source code. Impact analysis is required to optimize the testing effort.
In this paper, we propose an evaluation technique to predict impact propagation. Based on \nbprojects open-source Java projects and 5 classical mutation operators, we create \nbmutants mutants and study how the error they introduce propagates. 
This evaluation technique enables us to analyze impact prediction based on four types of call graph.
Our results show that graph sophistication increases the completeness of impact prediction.
However, and surprisingly to us, the most basic call graph gives the best trade-off between precision and recall for impact prediction.

\end{abstract}

\section{Introduction}
\label{sec:intro}

Software continuously evolves through changes affecting a module, a class, a function, \ldots~
A single change can impact the entire software package and may break many parts beyond the changed element, \eg a change in a file reading function can impact any class that uses it to read files.
When modifying the source code, developers think at whether the change will introduce an error and whether the error will propagate and impact other modules.
For this reason, many researchers have proposed techniques to reason on the impact of a given change (see \cite{li_survey_2013} for a recent survey).

The canonical problem statement of change impact analysis is: given a source code element $x$, what are the other source code elements impacted if one changes $x$?
In this paper, we consider this classical problem of impact analysis for object-oriented Java programs, where the granularity of analysis is the method: given a method $m$, what are the other methods impacted if one changes $m$? In essence, this is a prediction problem.

As all prediction problems, there is a trade-off between different dimensions \cite{lehnert_taxonomy_2011}.
First, whether the prediction is precise (predicted nodes are actually impacted by the change): this is known as the precision of the method.
Second, whether the prediction is complete (all actually impacted  nodes are predicted): this is known as the recall of the method.
Third, the time it takes to make the prediction. 
At the boundaries of the trade-off space, there are very precise systems, but they are slow and do not scale \cite{acharya_practical_2012}.
There are also very fast systems with very low precision. 
Finally, one can build degenerated cases predicting all nodes: these are complete by construction, but meaningless.

In this paper, we propose an evaluation technique based on mutation analysis intended to study these three dimensions of impact prediction. 
We use our evaluation technique for computing the accuracy of a change impact analysis based on call graphs.
Call graphs encode how methods and functions call each others. 
There are many different call graphs, depending on whether they are statically or dynamically extracted, whether the analysis is context-sensitive or not, \textit{etc}.
In this paper, we consider four different call graphs: one generated by JavaPDG \cite{shu_javapdg:_2013}, a publicly available tool, and three other call graphs that take into account different vectors of impact propagation, namely class fields and polymorphism.
Our main research question is: how do these four different graphs perform in terms of precision, completeness and execution time?

To answer this question, we present a novel experimental protocol, inspired from mutation testing.
We consider software equipped with a set of test cases (\aka a test suite), and we introduce arbitrary changes (mutations) in it. 
When running the test suite, some of the test cases fail; we consider the set of such failing test cases as being the ground truth that we have to predict.
To predict the impact of a given mutant in a method $m$, we compute the inverse transitive closure from $m$ according to a call graph and select only these nodes that are test cases.
To assess the performances of the technique, we compare the ground truth with the prediction. We obtain a confusion matrix from which we compute the precision and the recall of each call graph.

We run our protocol on \nbprojects mainstream open-source Java software packages.
For each of them, we create up to 3000 mutants using 5 different mutation operators.
Then we compare the precision and recall of the prediction depending on the call graph that is used.
Our results show that the sophistication indeed increases the completeness of impact prediction (higher recall).
However, and surprisingly to us, the simplest call graph gives the highest trade-off between precision and recall for impact prediction (as computed by the F-Score).

To sum up, our contributions are:
\begin{itemize}
	\item an algorithm to numerically analyze the accuracy of an impact analysis technique based on mutation testing;
	\item the definition of four kinds of call graphs for impact prediction.
	\item a large scale impact prediction experiment on \nbprojects projects and \nbmutants mutants comparing these 4 kinds of call graphs.
\end{itemize}

The remainder of this paper is structured as follows. 
Section \ref{sec:contribution} defines our protocol, 
Section \ref{sec:expres} presents our experiments and results, 
Section~\ref{sec:relworks} discusses the related work and 
Section \ref{sec:conclusion} concludes this paper.

\section{Contributions}
\label{sec:contribution}

In this section, we present our contributions. 
The first contribution is an evaluation technique of impact prediction (Section~\ref{sec:contrib1}). 
The second contribution is the definition and evaluation of four types of call graphs used for impact prediction (Section~\ref{sec:contrib2}). 
In addition, we propose a visual representation of the impact graph (Section \ref{section:visu}). 
Finally, we present the implementation of this work (Section \ref{sec:implem}).

\subsection{Definitions}
\label{sec:impact}

Change Impact Analysis (\aka CIA) is defined by Bohner \cite{bohner_software_2002} as ``the determination of potential effects to a subject system resulting from a proposed software change''. In this paper, we use Bohner's definition of the basic software change impact analysis process \cite{bohner_software_2002}. 

Assuming a change has been performed, Bohner defines the following sets used in impact analysis: 
\begin{inparaenum}[(i)]
	\item the ``starting impact set'' ($SIS$) is the list of software parts which can be impacted by the change;
	\item the ``candidate impact set'' ($CIS \subset SIS$) \footnote{also called the \emph{``estimated impact set'' (EIS)} in \cite{arnold_impact_1993}.} is the list of software parts predicted as impacted by a change impact analysis technique;
	\item the ``actual impact set'' ($AIS \subset SIS$) is the list of parts of the software which are actually impacted by the change;
	\item the ``false negative impact set'' ($FNIS$) is the list of missed impacts by the technique \footnote{Bohner named this set the ``discovered impact set'' ($DIS$), but this naming is not appropriate in our context and may be confusing.};
	\item the ``false positive impact set'' ($FPIS$) is the list of over-estimated impacts returned by the technique (\ie false positives).
\end{inparaenum}
Formally, the $FPIS$ and $FNIS$ sets are defined as:
	\begin{equation}
	FPIS = CIS-(AIS\cap{}CIS)
	\end{equation}
	\begin{equation}
	FNIS = AIS-(AIS\cap{}CIS)
	\end{equation}

\subsection{A Novel Evaluation Technique for CIA}
\label{sec:contrib1}

We present here a novel approach for evaluating a change impact analysis technique $I$.
The evaluation is based on the concept of actual impact set and candidate impact set presented in Section~\ref{sec:impact}. 
The closer the candidate impact set determined by $I$ is, the more accurate is the technique.

Our evaluation consists in assessing repetitively the impact prediction technique $I$ with a changed version of a program to determine how accurate the prediction is.
These changed versions of the programs are artificially obtained using mutation testing, as presented in Section~\ref{sec:contrib1a}.
Running the test cases on mutants produces the actual impact set of failing tests.

In Section~\ref{section:globalmetrics}, a way of computing the accuracy of $I$ is presented.
It relies on four metrics: the precision, the recall, the $F$-score and the completeness.

\subsubsection{Impact Evaluation with Mutants}
\label{sec:contrib1a}

\SetKwFunction{mutate}{mutants}
\SetKwFunction{findfailingtests}{failingTests}
\SetKwFunction{propnodes}{impactedTests}
\SetKwFunction{getestcases}{testCases}
\SetKwFunction{getcodeelement}{filterElements}

\newcommand{\soft}{\Sigma}
\newcommand{\testcases}{T}
\newcommand{\softgraph}{G}
\newcommand{\mutantoperator}{m_{op}}
\newcommand{\softwarelement}{e}
\newcommand{\mutant}{m}

\begin{algorithm}[t]
	\caption{Computes the candidate and actual impacted sets using mutation injection, test execution and call graph.}
	\label{code:algo_sets}
  \KwIn{$\soft$ the software package. $I$ an impact prediction technique. $\mutantoperator$ a mutation operator.}
  \KwOut{
    a map containing for each mutant (key) the CIS and AIS sets.
  }
  \Begin{
  	$IP \gets empty\_map()$\\
  	$\testcases \leftarrow \getestcases(\soft)$\label{algo:line2}\\
   \For{each $\softwarelement$ in $\getcodeelement(\soft, \mutantoperator)$\label{algo_sets:line3}}{
      \For{each $\mutant$ in $\mutate(\soft, \softwarelement, \mutantoperator)$\label{algo_sets:line4}}{
         \If{$\mutant$ compiles and is killed\label{algo_sets:line5}}{
           $CIS_m \leftarrow \propnodes(\mutant, I)$\label{algo_sets:line6}\\
           $AIS_m \leftarrow \findfailingtests(\mutant, \testcases)$\label{algo_sets:line7}\\
           $IP_m \gets \{AIS_m, CIS_m\}$
       }
     }
  }
  
  \Return{$IP$}
}
\end{algorithm}

We use mutation testing to obtain data used for determining the accuracy of a change impact analysis technique. 
Software mutants are used here as a way to simulate artificial faults. 
Indeed, a mutation consists in a random change in the source code. 
This is likely to result in an unexpected (\ie faulty) behavior, and thus in failing test cases. 
These failing test cases are the actual impact set. 
Then, using a change impact technique $I$, the candidate impact set is obtained.

Algorithm~\ref{code:algo_sets} illustrates the global process of generating changes, obtaining the actual impacts (\ie the $AIS$ -- Actual Impacted Set) and the estimated impact (\ie the $CIS$ -- Candidate Impacted Set) using an impact prediction technique $I$. This algorithm takes as input:
\begin{inparaenum}[(i)]
  \item the software package under study, 
  \item an impact prediction technique, and
  \item a mutation operator that is responsible for mutation injection. 
\end{inparaenum}

The output of the algorithm is a map which contains for each mutant, the set of actual impact set (AIS) and the candidate impact set (CIS).
In line \ref{algo:line2}, we get the set of test cases ($\getestcases$) from the input software $\soft$. In lines \ref{algo_sets:line3}--\ref{algo_sets:line5}, we select ($\getcodeelement$), mutate ($\mutate$) and test the appropriate elements in the software.
Appropriate elements are syntactic entities to which the specific change can be applied. 
In line \ref{algo_sets:line6}, we determine the test cases impacted by the mutation ($\propnodes$) according to the impact prediction technique $I$ (\ie the candidate impacted set).
In line \ref{algo_sets:line7}, the function $\findfailingtests$ returns the set of test cases that fail when running the mutated version of the software (\ie the actual impacted set). 

Some mutants are said to be unbounded. 
An \emph{unbounded mutant} is a mutant for which an impact prediction technique is not capable of predicting something because of a lack of information (which is different from predicting no impact).
The reason for which this happens is related to the prediction technique under consideration. 
For the one considered in this paper, the unbounded mutants are discussed in Section~\ref{sec:contrib2}.

\subsubsection{One-Impact Mutant-Level Accuracy Metrics}
\label{section:metrics}

In this section, we define 3 metrics used to analyze the output of Algorithm~\ref{code:algo_sets} for each mutant (\ie each prediction). These 3 metrics quantify and characterize the accuracy of an error impact analysis.

The \emph{precision} $P$ is the proportion of test cases predicted by the impact prediction technique which are actually impacted. It is computed using Equation~(\ref{eq:precision}).
The \emph{recall} $R$ is the proportion of test cases predicted by the call graph with regards to all test cases that are actually impacted. It is computed using Equation~(\ref{eq:recall}). 
The \emph{F-score} $F$ combines both metrics by computing their harmonic mean as in Equation~(\ref{eq:fscore}).
The precision, recall, and F-score are computed for a given mutant $m$.
We have:
\begin{equation}
\label{eq:precision}
P_m = \frac{|AIS_m \cap CIS_m|}{|CIS_m|},
\end{equation}
\begin{equation}
\label{eq:recall}
R_m = \frac{|AIS_m \cap CIS_m|}{|AIS_m|},
\end{equation}
\begin{equation}
\label{eq:fscore}
F_m = 2 \times \frac{P_m \times R_m}{P_m + R_m},
\end{equation}

where vertical bars such as $|E|$ denote the cardinality of the set E.

\subsubsection{Global Accuracy Metrics}
\label{section:globalmetrics}

In this section, we present the metrics used to determine the accuracy of a change impact analysis technique $I$ as a whole.
This is a global accuracy over observations made on results over all impacts presented in Section~\ref{section:metrics}.

Let $\K$ be the set of all killed mutants considered in a given experiment. A mutant is considered as killed as soon as at least one test case fails after running the mutant, while it did not fail on the un-mutated version of the program.
The accuracy of a change impact prediction technique is characterized by the average of the precision ($P$), the recall ($R$) and the $F$-scores ($F$) over all elements of $\K$.

Moreover, inspired by Arnold \etal \cite{arnold_impact_1993}, we define four sets to categorize the four types of possible predictions: $\S$ (\emph{same}), $\O$ (\emph{overestimate}), $\U$ (\emph{underestimate}) and $\D$ (\emph{different}).
These are based on Bohner sets presented in Section~\ref{sec:impact}.
Each mutant belongs to either one of these 4 sets. 
For a given mutant, we compute $FPIS$ and $FNIS$. Then, there are four cases:
\begin{itemize}
  \item if $FPIS = FNIS = \emptyset$, the mutant belongs to the $\S$ set. It implies that the $CIS$ and the $AIS$ are strictly equal ($AIS\cap{}CIS=AIS=CIS$)). \newline $\frac{|\S|}{|\K|}$ is the proportion of cases for which our method finds all and only actual impacts, which implies we cannot do better predictions for these cases;
  \item if $FPIS \ne \emptyset$ and $FNIS = \emptyset$, the mutant belongs to the $\O$ set. In this case, we have $AIS \subset CIS$. The change impact analysis technique is able to determine all impacts but it over-estimates them as it returns more impacts than actually happens. These scenarios are not perfect but are considered as safe \cite{arnold_impact_1993} as they return at least all the impacted elements;
  \item if $FPIS = \emptyset$ and $FNIS \ne \emptyset$, then the mutant belongs to the $\U$ set. In this case, we have $CIS \subset AIS $. The change impact analysis technique under-estimates the impact set as it returns less elements than the number of elements actually impacted;
  \item if $FPIS \ne \emptyset$ and $FNIS \ne \emptyset$, then the mutant belongs to the $\D$ set. The change impact analysis technique returns different impacts than the actual ones (even if some impacts may be estimated correctly).
\end{itemize}

In the two last cases, the change impact analysis technique under study misses impact candidates.
These 4 sets are disjoint, and each killed mutant belongs to either one of these 4 sets. $\{ \S, \O, \U, \D \}$ is a partition of the set of killed mutants $\K$; hence, we have:

\begin{equation}
  |\S| + |\O| + |\U| + |\D| = |\K|
\end{equation}

Algorithm \ref{code:arnold_sets} describes how each mutant is assigned to a set.

\begin{algorithm}[t]
	\caption{Computes the sets $\S$, $\O$, $\U$ and $\D$ for a set of mutants and their actual and candidate impacted sets.}
	\label{code:arnold_sets}
  \KwIn{$IP$ the map containing each mutant and its actual and candidate impacted sets (obtained using Algorithm~\ref{code:algo_sets})}
  \KwOut{
	$\S$, $\O$, $\U$ and $\D$: sets of mutants as defined in the text.
  }
  \Begin{
      $\S \gets \O \gets \U \gets \D \gets \emptyset$\\
  	\For{each $\mutant$ in IP}{
	  	$AIS, CIS \gets IP_\mutant$\\
	  	$FPIS \gets CIS-(AIS\cap{}CIS)$\\
	  	$FNIS \gets AIS-(AIS\cap{}CIS)$\\
	  	
	  	\uIf{$FPIS = \emptyset$ and $FNIS = \emptyset$}{
	  		$\S \leftarrow \S \cup \{\mutant\}$
	  	}\uElseIf{$FPIS \ne \emptyset$ and $FNIS = \emptyset$}{
	  		$\O \leftarrow \O \cup \{\mutant\}$
	  	}\uElseIf{$FPIS = \emptyset$ and $FNIS \ne \emptyset$}{
	  		$\U \leftarrow \U \cup \{\mutant\}$
	  	}\Else{
	  		$\D \leftarrow \D \cup \{\mutant\}$
	  	}
	}
	
	\Return{$\S$, $\O$, $\U$, $\D$}
}
\end{algorithm}

We also define the set $\C$ (\emph{complete}) as being the set of mutants for which the candidate impact set contains all actually impacted methods, maybe more. In other words, for these mutants, the change impact analysis method does not miss any impact. Formally, we define the set $\C$ as:

\begin{equation}
  \C = \S \cup \O
\end{equation}

We define the completeness as $p_\C = \frac{|\C|}{|\K|}$. It quantifies the extent to which a given call graph approximates the impact of a given mutation.
$p_\S = \frac{|\S|}{|\K|}$ quantifies the extent to which a given call graph perfectly determines the impact of a given mutation.

Unbounded mutants presented in Section~\ref{sec:contrib1a} belong to the $\U$ set ($\N \subset \U$).
The precision and recall for these unbound mutants are both equal to 0.

\subsection{Call Graphs for Impact Prediction}
\label{sec:contrib2}

Now, we present a family of four different types of call graphs we use for impact prediction. 
Each member of this family abstracts a particular way error may propagate in a piece of software.
A discussion about the reason we choose to use call graphs as a change impact prediction technique is presented in Section~\ref{sec:othercg}.
\emph{To the best of our knowledge, no author has proposed to use such variants of call graphs from the viewpoint of change impact prediction. Consequently, no accuracy comparison study of these call graphs has ever been made before.}

\begin{figure}
	\centering
	\includegraphics[width=10cm]{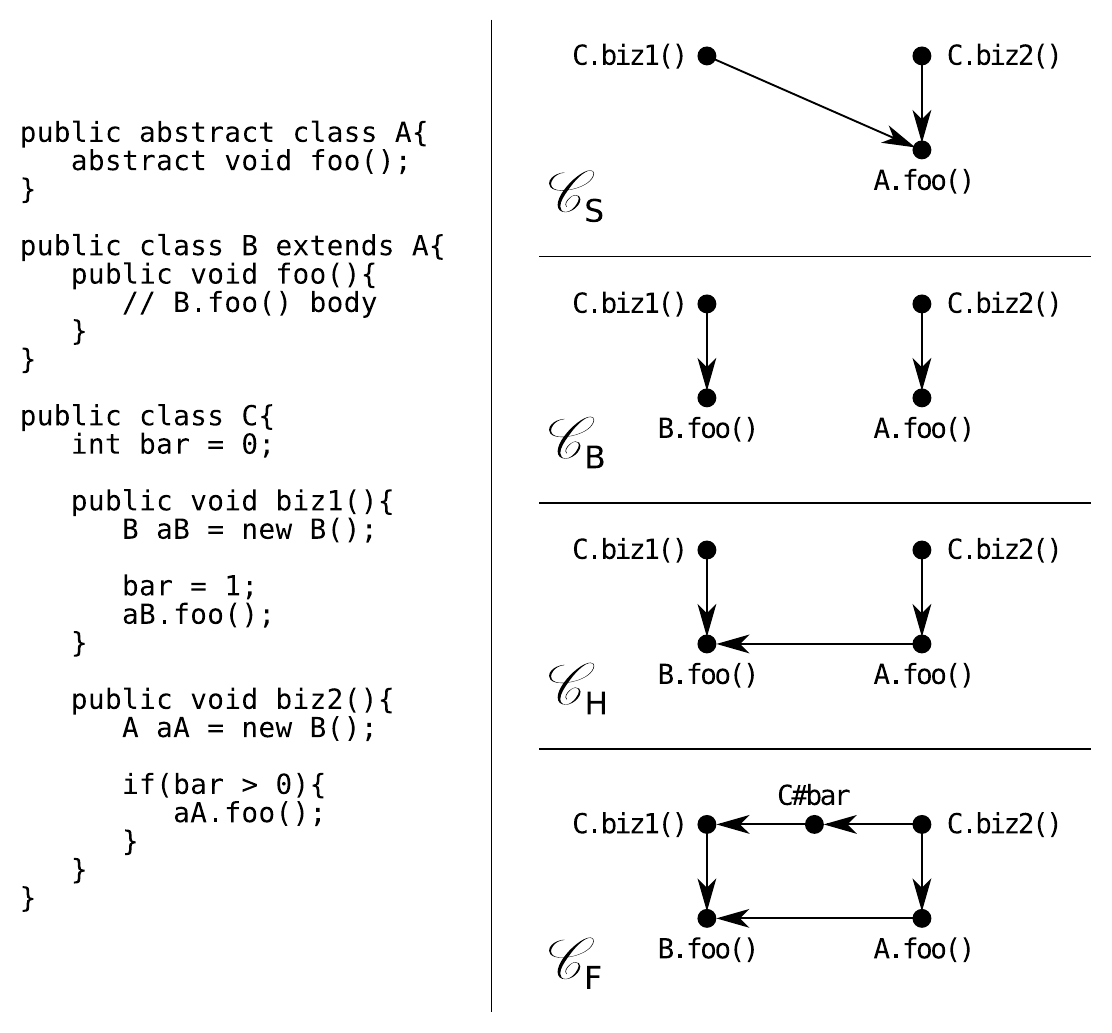}
	\caption{Simple Java source code and the four types of graphs obtained from it: \cgj, \cg, \cgcha and \cgf.}
	\label{fig:cgs}
\end{figure}

Call graphs model how software methods are called. If an error is present in a software method, methods calling it may themselves be impacted by the error. 
Exploring the call graph is a way for estimating the impact of a change. 
As an example, a \texttt{drawSquare} method calls a \texttt{drawLine} one. 
In the resulting call graph, there is an edge such as \texttt{drawSquare} $\longrightarrow$ \texttt{drawLine}. 
If the \texttt{drawLine} method has been changed, this is likely that the \texttt{drawSquare} method which calls it (\ie depends on it) will be also impacted by the change.
In this paper, we take the definition of call graph by Grove \etal \cite{grove_call_1997}: ``\emph{the program call graph [as] a directed graph that represents the calling relationships between the program’s procedures (...) each procedure is represented by a single node in the graph. Each node has an indexed set of call sites, and each call site is the source of zero or more edges to other nodes, representing possible callees of that site}''. 
However, this definition of call graph allows many variations. Hence, we consider in this paper a family of 4 different call graphs. Table~\ref{table:usegraphs} lists them. Figure~\ref{fig:cgs} illustrates the key differences between these 4 call graphs.

\begin{table*}
  \caption{The four types of call graph we define for error impact prediction.}\label{table:usegraphs}
  \centering
  \small
  \begin{tabular}{@{}l c c p{8cm} @{}}
    \toprule
	Name & Hierarchy & Fields & Description \\
    \midrule
    \cgj & No & No & Call graph extracted using JavaPDG \cite{shu_javapdg:_2013}.\\
    \cg & No & No & Call graph extracted using softminer considering only method calls. Calls to inherited methods are resolved.\\
    \cgcha & Yes & No & \cg with Class Hierarchy Analysis (CHA), a standard call graph in object-oriented static analysis.\\
    \cgf & Yes & Yes & \cgcha with field analysis: each read/write access to a field may propagate an error.\\
    \bottomrule
  \end{tabular}
\end{table*}

The first one is the call graph obtained using the JavaPDG tool by Shu \etal \cite{shu_javapdg:_2013}. We refer to such a call graph as \cgj, where ''S'' refers to the first author of the paper. In such a call graph, overriding methods are not resolved. Thus, if the method \texttt{A.foo()} overrides the method \texttt{B.foo()}, and the method \texttt{C.bar()} calls \texttt{A.foo()}, the call graph will contain a call from \texttt{C.bar()} to \texttt{A.foo()}. Figure~\ref{fig:cgs} gives another example of this point: methods \texttt{biz1()} and \texttt{biz2()} both call the same method, but the former calls it on a \texttt{B} object and the latter on an \texttt{A} object. However, in the call graph, both are resolved with the same node.
	
\cg is a similar basic call graph which uses the signature of the class according to the static type of the receiver, as illustrated in Figure~\ref{fig:cgs}. We see that in this call graph, the method \texttt{biz1()} and \texttt{biz2()} both call a \texttt{foo()} method, but the former on a \texttt{B} object and the latter on a \texttt{A} object. Formally, for \cgj and \cg, if method \texttt{m} calls method \texttt{n}, there is an edge $node_m \longrightarrow node_n$.
However, errors may propagate through edges that are neither in \cgj nor in \cg. 
Thus, we propose two other flavors enriching \cg by handling some object-oriented programming concepts.
	
\cgcha takes into consideration the class hierarchy analysis (\aka CHA)\cite{dean_optimization_1995} to take inheritance and interface implementation into consideration. To do so, for each method, we explore the classes extended and the interfaces implemented by the class in which the method is defined. We add edges from the parent definition method to the overridden method in the hierarchy. Formally, if a method \texttt{m} implements an abstract class or an interface method \texttt{n}, there is an edge $node_n \longrightarrow node_m$. This is illustrated on Figure~\ref{fig:cgs} where we observe that an edge has been added from \texttt{A.foo()} to \texttt{B.foo()}.
	
\cgf takes into consideration CHA but also reads and writes to fields. Indeed, when a method writes to a field, it modifies its content and thus, potentially inserts an error in it. In the opposite situation, a method which reads a variable on which an error has been inserted may be impacted by this error.
Thus, when writing to a variable, the propagation goes from the method to the variable, but on the opposite way, when reading, the propagation goes from the variable to the method. Formally, if method \texttt{m} reads the field \texttt{f}, there is an edge $node_m \longrightarrow node_f$. If \texttt{m} writes the field \texttt{f}, there is an edge $node_m \longleftarrow node_f$. This is illustrated on Figure~\ref{fig:cgs}: a node has been added for the \texttt{bar} field and two edges have been added: one from \texttt{C.biz2()} to \texttt{C\#bar} for the write operation and one from \texttt{C\#bar} to \texttt{C.biz1()} for the read operation. This feature is similar to the method-level data dependency edge presented by Shu \etal \cite{shu_mfl:_2013} with the difference that we add a node and two edges between the calls where they directly add an edge. However, from a propagation point-of-view, both approaches are totally equivalent.

\subsubsection{Elaborated Example}
\label{sec:cgeval}

\begin{figure}[t]
  \centering
  \includegraphics[width=80mm]{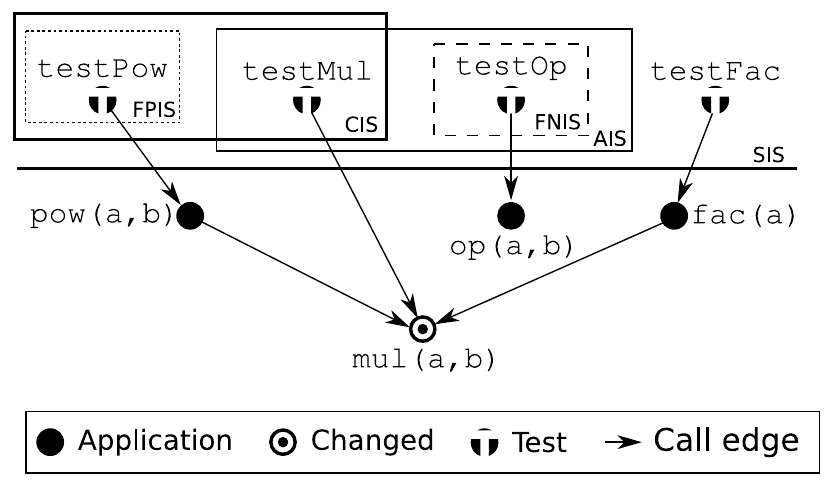}
  \caption{Example of a call graph in which a change has been introduced. The class graph includes application nodes, test nodes and call edges. The rectangles illustrate Bohner's sets.}
  \label{fig:graph}
\end{figure}

Figure~\ref{fig:graph} illustrates a simple application of the Algorithm~\ref{code:algo_sets} using a call graph. Three types of nodes are presented: application nodes (plain circle), test nodes (circle with a T) and the changed node which is itself an application node (double circle). The \texttt{mul} method is a multiplication method. As we can see, both the power method (\texttt{pow}) and the factorial method (\texttt{fac}) use the multiplication one. Moreover, another operand method (\texttt{op}) is also defined but not called explicitly in the call graph. This method uses reflection (which is not resolved statically) to call the \texttt{mul} method, resulting in the absence of edge between \texttt{op} and \texttt{mul}. Moreover, each method has its associated test method prefixed by \texttt{test}.
	
\label{sec:arnold_sets}
All test nodes belong to the \emph{System Impacted Set (SIS)} and are all potentially impacted. We use mutation injection to produce a change to the \texttt{mul} method. Running test cases on the changed version of the code gives a list of failing and passing test cases. As these results are obtained by the execution of the program, the failing test cases make the actual impacted set (AIS). In this example, we suppose that there are two actually impacted test cases: \texttt{testMul} and \texttt{testOp} illustrated by the thin box.
	
As explained earlier, we use call graphs as impact analysis technique. In our example, we determine which nodes are connected to the impacted one (the \texttt{mul} method node). By exploring recursively the edges in the reverse direction, we reach two test nodes: \texttt{testPow} and \texttt{testMul}. These form the candidate impacted set ($CIS$), illustrated by the thick box.
	
Two other sets can also be observed. The \texttt{testPow} test method is a false positive as it is reported as impacted by our impact prediction technique but does not actually fail when running the test cases program. These test cases belong to the \emph{False Positives Impacted Set (FPIS)} illustrated by a dotted box.  On the other hand, the \texttt{testOp} test method is a false negative: running the test cases reports this test method as impacted, but there is no path from the impacted method (\texttt{mul}) to the \texttt{testOp} test method. These test cases belong to the \emph{False Negatives Impacted Set (FNIS)} illustrated by a dashed box.

Let us now discuss the case of unbounded mutants $\N$ ($\N \subset \U$) presented in Section~\ref{sec:contrib1a}. $\N$ contains all mutants for which the prediction is not possible because of the call graph.
This happens for different reasons: certain call graphs such as the \cgj may contain only nodes corresponding to the first definition of a method and do not resolve the inherited ones. Thus if the change occurs in an overridden method, it would not be found in the call graph. Another scenario is when the mutation occurs in a method which is defined but not actually called in the code (\eg as \texttt{equals}). 
Mutants in $\N$ set are removed from $\K$. Clearly, the set of mutants belonging to $\N$ depends on the call graph being used. This point is visible in the experimental section, where we give the cardinality of $\N$ for each type of call graph we work with.

\subsection{Visualization}
\label{section:visu}

\begin{figure}[t]
	\centering 
	\ifthenelse{\equal{\figureinlatex}{true}}{
	\def\svgwidth{80mm} 
\begingroup
  \makeatletter
  \providecommand\color[2][]{
    \errmessage{(Inkscape) Color is used for the text in Inkscape, but the package 'color.sty' is not loaded}
    \renewcommand\color[2][]{}
  }
  \providecommand\transparent[1]{
    \errmessage{(Inkscape) Transparency is used (non-zero) for the text in Inkscape, but the package 'transparent.sty' is not loaded}
    \renewcommand\transparent[1]{}
  }
  \providecommand\rotatebox[2]{#2}
  \ifx\svgwidth\undefined
    \setlength{\unitlength}{252bp}
    \ifx\svgscale\undefined
      \relax
    \else
      \setlength{\unitlength}{\unitlength * \real{\svgscale}}
    \fi
  \else
    \setlength{\unitlength}{\svgwidth}
  \fi
  \global\let\svgwidth\undefined
  \global\let\svgscale\undefined
  \makeatother
  \begin{picture}(1,1.33333333)
    \put(0,0){\includegraphics[width=\unitlength,page=1]{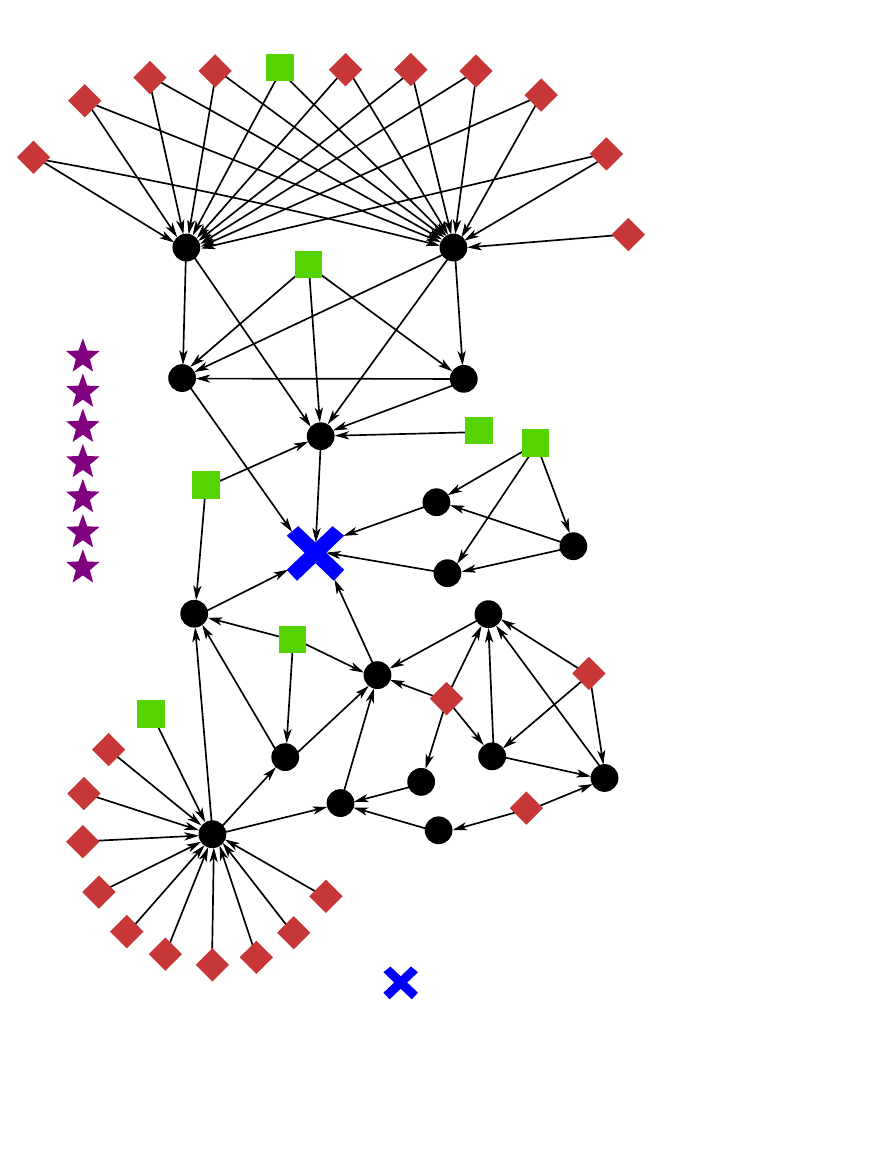}}
    \put(1.7417589,2.8536232){\color[rgb]{0,0,0}\makebox(0,0)[lb]{\smash{}}}
    \put(1.45561281,2.71004396){\color[rgb]{0,0,0}\makebox(0,0)[lb]{\smash{}}}
    \put(2.28350578,2.70523371){\color[rgb]{0,0,0}\makebox(0,0)[lb]{\smash{}}}
    \put(0,0){\includegraphics[width=\unitlength,page=2]{fig/apacheCommonsLang_if_mut107v2.pdf}}
    \put(0.49256072,0.19933599){\color[rgb]{0,0,0}\makebox(0,0)[lb]{\smash{Introduced error}}}
    \put(0,0){\includegraphics[width=\unitlength,page=3]{fig/apacheCommonsLang_if_mut107v2.pdf}}
    \put(0.49256072,0.15958716){\color[rgb]{0,0,0}\makebox(0,0)[lb]{\smash{False Positives ($FPIS$)}}}
    \put(0,0){\includegraphics[width=\unitlength,page=4]{fig/apacheCommonsLang_if_mut107v2.pdf}}
    \put(0.49545211,0.12415125){\color[rgb]{0,0,0}\makebox(0,0)[lb]{\smash{True Positives ($CIS \cap AIS$)}}}
    \put(0,0){\includegraphics[width=\unitlength,page=5]{fig/apacheCommonsLang_if_mut107v2.pdf}}
    \put(0.49256072,0.08871535){\color[rgb]{0,0,0}\makebox(0,0)[lb]{\smash{False Negatives ($FNIS$)}}}
    \put(0,0){\includegraphics[width=\unitlength,page=6]{fig/apacheCommonsLang_if_mut107v2.pdf}}
    \put(0.49514482,0.05110042){\color[rgb]{0,0,0}\makebox(0,0)[lb]{\smash{Application nodes}}}
  \end{picture}
\endgroup

	}{
	\includegraphics[width=80mm]{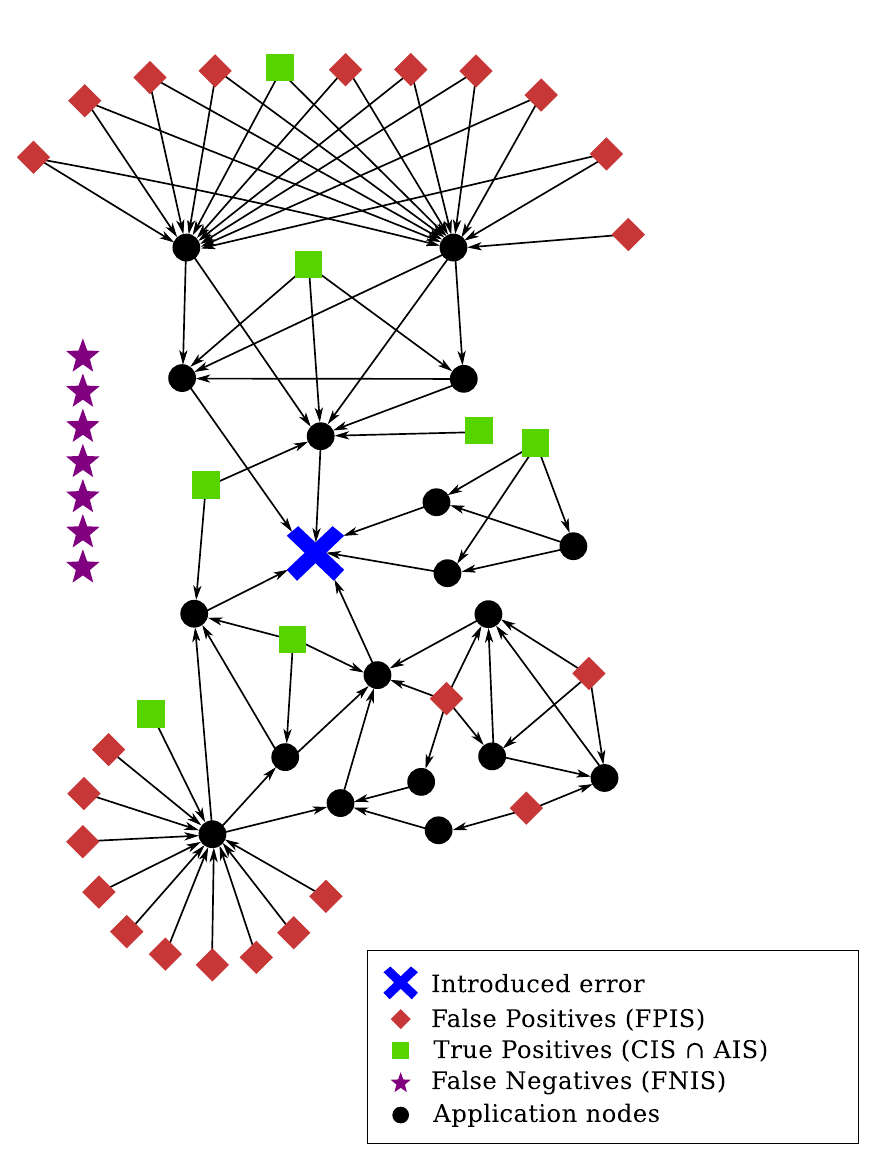}
	}
	\caption{Effect of a particular mutation in the Apache Commons Lang project. Only the interesting part of the call graph is represented here; the call graph is much larger, made of 6000+ nodes. Black circle are the nodes that propagate the mutation injected in the node denoted with a blue cross. Nodes illustrated by green boxes, red diamonds and purple stars are test cases related to the injected mutation. Green boxes nodes are test cases that are correctly predicted as impacted by the injected mutation; these are true positives. Red diamonds nodes are test cases that are predicted as impacted, but are not; these are false positives. Purple stars nodes are test cases that should have been predicted as being impacted but have not been; these are false negatives.} 
	\label{fig:impactex}
\end{figure}

In this section, we propose a visualization of error propagation: this feature provides developers with an idea of the potential consequences of software mutation and the complexity of its impact.
Figure \ref{fig:impactex} illustrates an error-introducing change in Apache Commons Lang. 
This illustration includes concepts of both contributions: the call graph \textit{per se} and the impacted sets presented in the evaluation technique.
Each node represents either a method or a field, and each edge represents a call to a method.
The blue cross is the node where the mutation occurs.
Purple stars are missed test cases (\ie detected only by the test suite execution),
red diamonds are incorrectly predicted test cases (\ie predicted by the call graph but not by the test suite execution),
green boxes are correctly predicted test cases (\ie found by both techniques) and 
black circles are application nodes. As an example, the graph illustrated on Figure \ref{fig:impactex} is composed of 56 nodes with 7 correctly predicted, 7 missed test cases, 23 incorrectly predicted nodes and 19 application nodes. As there are missed and incorrectly predicted test cases, this error-introducing change belongs to the $\D$ set.
In the example, we notice the multiple propagation paths that exist from the node at which the error has been introduced to the impacted test nodes.

\subsection{Implementation}
\label{sec:implem}

	The experiments conducted in this paper are implemented in three different tools. The code of these tools is publicly available on Github\footnote{\url{https://github.com/v-m/PropagationAnalysis}. The version used for extracting graphs and running the experiments of this paper is version tag \texttt{g1}.}.
	The first tool is named \emph{simple mutation framework (smf)}: this is our mutation tool. Several mutation tools exist, for instance Javalanche\footnote{\url{http://javalanche.org}}, or Pitest\footnote{\url{http://pitest.org}}. However, we need a full control over the mutation process and on extracted information. So, we have implemented ours.
	The second tool, named \emph{softminer}, extracts the call graph from Java source code.
	The third tool is named \emph{pminer}: it implements the prediction analysis (Algorithms \ref{code:algo_sets} and \ref{code:arnold_sets}, propagation prediction from call graphs and accuracy computations). 
	The call graph being used is either generated by \emph{softminer}, or generated by \emph{JavaPDG}. Both \emph{smf} and \emph{softminer} use \emph{Spoon}, an open-source library for analyzing and transforming Java source code \cite{pawlak_spoon:_2015}.

\begin{table*}
  \caption{List of mutation operators considered in this paper. Java operators $T$ and $F$ stand respectively for $true$ and $false$ boolean types. With binary operators, $L$ and $R$ stand respectively for \emph{left operand} and \emph{right operand}.}\label{table:mutop}
  \centering
  \small
  \begin{tabular}{@{}l l l@{}}
    \toprule
    ID & Name & Java operators \\
    \midrule
    ABS & Absolute value insertion & $java.lang.Math.abs()$ \\
    AOR & Arithmetic operator replacement & $+$, $-$, $*$, $/$, $\%$, $L$, $R$ \\
    LCR & Logical connector replacement & $\&\&$, $||$, $T$, $F$, $L$, $R$ \\
    ROR & Relational operator replacement & $<$, $<=$, $>$, $>=$, $==$, $!=$, $T$, $F$ \\
    UOI & Unary operator inversion & $!$, $++$, $--$ \\
    \bottomrule
  \end{tabular}
\end{table*}

Our technique requires mutation operators. 
We consider the five mutation operators presented by King and Offutt \cite{king_fortran_1991}. As shown by Offutt \etal, these five operators are sufficient to effectively implement mutation testing \cite{offutt_experimental_1996}. These operators are listed in Table~\ref{table:mutop}: the leftmost column is the three letter acronym used by King and Offutt, the central column is the full name, and the rightmost column lists the set of operators implied in the mutation. 
A mutation operator changes a single atomic element. Any software source code elements may be considered in a mutation.

As these operators are originally intended for the Fortran programming language, we adapted them in order to make them compatible with Java programming language (see the Java operators implied on rightmost column of Table~\ref{table:mutop}).
Our Fortran to Java adaptations of the operators are:
\begin{inparaenum}[(i)]
	\item \emph{Absolute value insertion (ABS)} in which each numerical expression (variable or method call) or literal is replaced by its absolute value.
	\item \emph{Arithmetic operator replacement (AOR)} in which each arithmetic expression using Java arithmetic operators \texttt{+}, \texttt{-}, \texttt{*}, \texttt{/}, \texttt{\%} is replaced by a new arithmetic expression with the same operands but where the operator is changed into another one of the same family, chosen uniformly at random. Two other mutation candidates are also the left and the right operand alone, after removing the operator and one of the two operands;
	\item \emph{Logical connector replacement (LCR)} in which each logical expression using Java logical operators \texttt{\&\&} and \texttt{||} is replaced by a new logical expression with the same operands but where the logical operator is changed by another one. Moreover, each logical expression may also be mutated by the constants \texttt{true} and \texttt{false}. Two other mutation candidates are also the left and the right operand alone, after removing the operator and one of the two operands.
	\item \emph{Relational operator replacement (ROR)} in which each relational expression using Java relational operators \texttt{<}, \texttt{<=}, \texttt{>}, \texttt{>=}, \texttt{==} and \texttt{!=} is mutated to a relational expression with the same operands but where the relational operator is changed with another one. Moreover, each relational expression may be replaced by the constants \texttt{true} and \texttt{false};
	\item \emph{Unary operator inversion (UOI)} in which each arithmetic and logical expression is mutated. Arithmetic expressions are mutated to their opposite value (\ie multiplied by \texttt{-1}), their incremented value (\ie add \texttt{1}) and their decremented value (\ie subtract \texttt{1}). Logical expressions are complemented (\ie apply  the \texttt{not} (\texttt{!}) Java operator).
\end{inparaenum}

\section{Experimental Evaluation}
\label{sec:expres}

We run our Algorithms \ref{code:algo_sets} and \ref{code:arnold_sets} for the 4 kinds of call graphs presented in Table \ref{table:usegraphs}, on a dataset of \nbprojects Java programs using 5 mutation operators. 
This enables us to answer the following research questions:

\newcommand{\rqperf}{What is the difference between the different types of call graphs in terms of impact prediction accuracy?}
\newcommand{\rqprojmutdep}{Is impact prediction project-dependent or mutation-dependent?}
\newcommand{\rqreasonbadaccuracy}{What are the reasons of the bad accuracy of impact prediction using call graphs?}
\newcommand{\rqtime}{What is the trade-off between the accuracy and the time needed to compute the impact prediction?}

\RQInline{\rqperf}
We determine the prediction capabilities offered by each call graph and whether field analysis and inheritance analysis improve or decrease the prediction of error propagation.

\RQInline{\rqprojmutdep} 
It may happen that one call graph is good for predicting the error propagation given a specific mutation operator. 
This is what we call mutation-dependent error impact prediction.
The same question may be raised regarding projects.
Answering this question allows us to determine the level of genericity of our approach.

\RQInline{\rqreasonbadaccuracy} To answer this question, we manually investigate some cases where the prediction is poor to better understand the reasons leading to a discrepancy between predictions and the actual execution of code.

\RQInline{\rqtime} 
Running the test suite is a good and precise way to know the actual impact of a change, but this requires important execution time.
On the other hand, a method based on call graphs is cheap in time but less precise in its prediction.
As explained above, it may over-estimate or under-estimate the actual propagation of a change. 
We want to better characterize the trade-off between accuracy and time needed for impact analysis.

\RQRESET{}

\subsection{Dataset}

\begin{table*}
  \caption{Statistics about the projects considered in this paper.}\label{table:dataset}
  \centering
  \begin{tabular}{@{}l l l r@{}}
    \toprule
    Project & Version & Commit & LOC \\
    \midrule
    
Codec & 1.11 & r1676715 & 17,531\\
Collections & 4.1 & r1610049 & 55,081\\
Gson & 2.3.2 & \#fcfd397 & 20,072\\
Io & 2.5 & r1684201 & 26,528 \\
Jgit & 4.1.0 & \#3c33d09 & 133,865\\
Jodatime & 2.8.1 & \#6da4053 & 85,000\\
Lang & 3.5 & \#6965455 & 67,509\\
Shindig & 2.5.3 & r1687149 & 15,710\\
Sonarqube & 5.2 & \#1385dd3 & 29,342\\
Spojo & 1.0.7 & \#8fb2194 & 3,371\\
\midrule

\textbf{Total} & & & \textbf{454,009}\\

    \bottomrule
  \end{tabular}
\end{table*}

We consider a dataset composed of \nbprojects \emph{Java} software packages.
It is composed of the following projects: \emph{Apache Commons Lang}, \emph{Apache Commons Collections}, \emph{Apache Commons Codec}, \emph{Apache Commons Io}, \emph{Google Gson}, \emph{Jgit}, \emph{Jodatime}, \emph{Apache Shindig}, \emph{Spojo} and \emph{Sonarqube}. When the project is made of several sub-projects, we consider only the main one. Tables \ref{table:dataset} and \ref{table:datasetg} report the key descriptive statistics about these projects. Table~\ref{table:dataset} gives the name, the version, the git commit-id (starting with \#) or the svn revision number (starting with 'r') and the number of lines of code (computed using \emph{cloc} \footnote{\url{http://cloc.sourceforge.net/}}) of the software being analyzed. Table~\ref{table:datasetg} describes the different call graphs under investigation. This table is made of four couples of columns which give the number of nodes and edges composing each call graph, namely \cgj (call graph obtained using JavaPDG tool), \cg (our basic call graph), \cgcha (our call graph with CHA) and \cgf (our call graph with CHA and fields).
The generated data used in this paper are publicly available on Github \footnote{\url{https://github.com/v-m/PropagationAnalysis-dataset}}.

We observe that the \cgj contains less nodes and edges than \cg, \cgcha and \cgf (excepted for \textit{Gson}, where the \cgj has more nodes than \cg and \cgcha and for \textit{Collections}, where the \cgj has more nodes than \cg).
This is due to the fact that \cgj does not resolve the inherited method name, which means that if a method \texttt{A.foo} calls a method \texttt{B.bar} which extends \texttt{C.bar}, the graph only contains calls to the super method \texttt{C.bar}.

Since we have the same number of nodes for \cg and \cgcha, this validates our implementation because, we just added calls between some classes belonging to the same hierarchy. These methods are already present in \cg, they are just called by the callee. In \cgcha, we add edges between methods belonging to the same hierarchy, (\ie overridden methods). The number of nodes and edges increases in \cgf because we introduce nodes and edges to reflect fields and their use (reads and writes).

\subsection{Results}

We now address the research questions introduced in Section~\ref{sec:expres}. In particular, we present the accuracy for error impact analysis obtained with the different types of call graphs.

\RQ{\rqperf}
\label{rq:perf}

To answer this question, we compute the metrics presented in Section~\ref{section:globalmetrics}.
Their values are given in Table \ref{table:metrics_for_graphs}.
The first, second and third columns give the project name, the mutation operator, and the number of killed mutants for the project. The remainder of the table is split into four parts, one for each type of call graphs. In each part, the first column ($|\N|$) shows the number of mutants for which there is no node in the graph which corresponds to the method being mutated, or for which the corresponding node has no neighbor, \ie contains no in/out edges. The second column ($p_\S$) is the proportion of mutants for which the impact prediction is perfect (the failing test cases obtained from the call graph are exactly the ones obtained by test suite execution). The third column ($p_\C$) is the proportion of mutants for which the impact prediction is complete, \ie include all failing test cases. The fourth, fifth and sixth columns are the precision, recall and F-score averaged over all considered mutants. 
For each line, the value in bold font is the best F-score obtained among the four types of call graph.

\begin{table*}
	\caption{Statistics about the call graphs for the projects considered in this paper.}\label{table:datasetg}
	\centering
	\begin{tabular}{@{}l r r c r r c r r c r r@{}}
		\toprule
		& \multicolumn{2}{c}{\cgj} & & \multicolumn{2}{c}{\cg}&  & \multicolumn{2}{c}{\cgcha} &  & \multicolumn{2}{c}{\cgf} \\
		\cline{2-3}
		\cline{5-6}
		\cline{8-9}
		\cline{11-12}
		Project & \#N & \#E && \#N & \#E && \#N & \#E && \#N & \#E \\
		\midrule
		
		Codec & 1,338 & 1,959  && 1,490 & 2,218 && 1,490 & 2,336 && 1,884 & 3,588\\
		Collections & 6,008 & 7,747  && 6,678 & 9,252 && 6,678 & 12,047 && 7,637 & 17,178\\
		Gson & 2,630 & 5,492  && 2,480 & 5,381 && 2,480 & 5,674 && 3,317 & 9,101\\
		Io & 2,382 & 3,634  && 2,662 & 3,974 && 2,662 & 4,198 && 3,305 & 7,004\\
		Jgit & 11,571 & 31,647  && 12,560 & 35,953 && 12,560 & 37,679 && 17,350 & 60,458\\
		Jodatime & 8,531 & 23,283  && 9,809 & 31,329 && 9,809 & 33,991 && 11,879 & 44,956\\
		Lang & 6,033 & 8,892  && 6,220 & 9,004 && 6,220 & 9,345 && 7,577 & 16,094\\
		Shindig & 1,410 & 2,020  && 1,933 & 2,373 && 1,933 & 2,621 && 2,723 & 5,096\\
		Sonarqube & 3,126 & 5,025  && 4,322 & 5,737 && 4,322 & 5,852 && 5,960 & 10,706\\
		Spojo & 306 & 630  && 417 & 884 && 417 & 917 && 521 & 1,331\\
		\midrule
		
		\textbf{Total} & \textbf{43,335} & \textbf{90,329} && \textbf{48,571} & \textbf{106,105} && \textbf{48,571} & \textbf{114,660} && \textbf{62,153} & \textbf{175,512}\\
		
		\bottomrule
	\end{tabular}
\end{table*}

The first observation is that \cgj has an important number of unbound mutants, more than 50\% in some cases such as Codec with ABS mutation operator. 
The three other call graphs have less unbound mutants.
Further investigations show that the main reason of unbound mutants for \cgj is that the mutation occurs in an inherited node method which is not resolved by \cgj (as presented in Section~\ref{sec:cgeval}). For \cg, \cgcha and \cgf, unbound nodes are always nodes which are isolated (\ie no connected edges). We noticed that in \cg, \cgcha and \cgf, these nodes are not always called in the software. Examples of such methods are \texttt{equals}, \texttt{compare} or some state testing method such as \texttt{isInAlphabet} for Apache Commons Codec project.

This explains low scores for \cgj as every unbound mutants have a precision and a recall equal to 0. This strongly highlights the fact this graph is not complete enough to perform good impact prediction analysis.

Considering the F-score values (F), we see that \cg is the one which gives the best F-scores (in 18 cases out of 50, that is in more than 30\% of cases) which indicates it is the best suited call graph for impact prediction. However, \cgcha F-scores are close to the \cg ones: in 10 cases out of 50, \cgcha has similar values and in 16 cases out of 50 it has better ones, which means that \cgcha is also a good candidate for impact prediction. 

If a call graph shows better F-scores, the observation is valid for all mutation operators of the project, which indicates the impact prediction technique is project-dependent (see Research~Question~\ref{rq:projdep}).

Let us now consider the $p_\C$ metric, which indicates whether the prediction is sound (in which proportion of mutations impact that actually happens is not missed).
From \cgj to \cg, we have an average increase of $p_\C$ around 20\%. Then, considering fields and hierarchy indeed better captures the error propagation: the $p_\C$ metric increases in average of 15\% when taking into consideration class hierarchy analysis (from \cg to \cgcha) and of 5\% when also considering fields (\cgcha to \cgf).
However, if we look at the increase project by project, we see important differences. Considering the inclusion of hierarchy (from \cg to \cgcha), Gson and Jodatime have high $p_\C$ average increases of respectively 49\% and 41\%, which implies a high usage of hierarchy in these projects. At the opposite, Collections and Io have lower $p_\C$ values with both an average increase of 1.8\%. The $p_\C$ value can reach values as high as 100\% for Spojo. 
The best increases are for Lang (\resp Io) with AOR mutation operator where $p_\C$ raises from \cgj to \cg from 31\% to 90\% (\resp from 27\% to 86\%) and for Gson with ROR mutation operator where $p_\C$ raises from \cg to \cgcha from 36\% to 93\%.

\afterpage{
\thispagestyle{empty}
\begin{landscape}
	\begin{table}[htbp]
		\caption{The main metrics of impact prediction based on four different call graphs. $|\K|$ is the number of killed mutants. $|\N|$ is the number of unbound mutants (see Section~\ref{section:globalmetrics}). $p_\S=\frac{|\S|}{|\K|}$ is the proportion of mutants for which the predicted impact exactly matches the actual one. $p_\C=\frac{|\C|}{|\K|}$ is the proportion of mutants for which the prediction is complete  (\ie{} does not miss any impacted test case), $P$ is the precision averaged over all killed mutants, $R$ is the recall averaged over all killed mutants, and accordingly, $F$ is the $F$-score averaged over all killed mutants.}
		\label{table:metrics_for_graphs}
		\centering
		\tiny
\setlength{\tabcolsep}{4pt}
		\begin{tabular}{@{}lcrp{0.01mm}rrrrrrp{0.01mm}rrrrrrp{0.1mm}rrrrrrp{0.1mm}rrrrrr@{}}
			\toprule
			&&&& \multicolumn{6}{c}{\cgj} && \multicolumn{6}{c}{\cg} && \multicolumn{6}{c}{\cgcha} && \multicolumn{6}{c}{\cgf}\\
			\cmidrule{5-10}
			\cmidrule{12-17}
			\cmidrule{19-24}
			\cmidrule{26-31}
			Project & Ope. & $|\K|$ && $|\N|$ & $p_\S$ & $p_\C$ & P & R & F & & $|\N|$  & $p_\S$ & $p_\C$ & P & R & F & & $|\N|$ & $p_\S$ & $p_\C$ & P & R & F & & $|\N|$ & $p_\S$ & $p_\C$ & P & R & F\\
			\midrule
			Codec & ABS & 302  && 170 & 2\%& 26\%& 0.24 & 0.30 & 0.15 & & 5 & 3\%& 62\%& 0.43 & 0.68 & \textbf{0.31} & & 3 & 2\%& 79\%& 0.25 & 0.87 & 0.29 & & 0 & 1\%& 88\%& 0.18 & 0.93 & 0.21 \\
			& AOR & 458 &&  157 & 3\%& 40\%& 0.36 & 0.43 & 0.23 & & 6 & 3\%& 70\%& 0.42 & 0.75 & \textbf{0.34} & & 5 & 0\%& 81\%& 0.25 & 0.89 & 0.30 & & 0 & 0\%& 86\%& 0.19 & 0.92 & 0.23 \\
			& LCR & 497 &&  128 & 0\%& 43\%& 0.25 & 0.49 & 0.21 & & 9 & 2\%& 62\%& 0.36 & 0.68 & \textbf{0.28} & & 0 & 1\%& 83\%& 0.17 & 0.91 & 0.24 & & 0 & 1\%& 89\%& 0.15 & 0.95 & 0.21 \\
			& ROR & 484 &&  173 & 4\%& 32\%& 0.37 & 0.39 & 0.26 & & 14 & 5\%& 55\%& 0.51 & 0.65 & \textbf{0.39} & & 5 & 2\%& 78\%& 0.29 & 0.88 & 0.33 & & 0 & 2\%& 82\%& 0.24 & 0.90 & 0.26 \\
			& UOI & 470 &&  139 & 3\%& 42\%& 0.39 & 0.47 & 0.26 & & 6 & 4\%& 66\%& 0.48 & 0.73 & \textbf{0.39} & & 3 & 1\%& 79\%& 0.28 & 0.88 & 0.32 & & 0 & 1\%& 85\%& 0.23 & 0.91 & 0.26 \\
			\midrule
			Collections & ABS & 220 && 71 & 12\%& 20\%& 0.57 & 0.22 & 0.19 & & 13 & 25\%& 35\%& 0.84 & 0.42 & 0.38 & & 9 & 25\%& 37\%& 0.74 & 0.44 & \textbf{0.39} & & 0 & 18\%& 41\%& 0.51 & 0.49 & 0.30 \\
			& AOR & 380 && 194 & 10\%& 12\%& 0.43 & 0.13 & 0.12 & & 59 & 35\%& 43\%& 0.89 & 0.47 & \textbf{0.45} & & 15 & 34\%& 44\%& 0.77 & 0.48 & 0.44 & & 0 & 30\%& 49\%& 0.58 & 0.54 & 0.39 \\
			& LCR & 332 && 69 & 20\%& 39\%& 0.65 & 0.41 & 0.35 & & 23 & 28\%& 51\%& 0.81 & 0.56 & \textbf{0.48} & & 16 & 28\%& 52\%& 0.70 & 0.57 & 0.47 & & 3 & 22\%& 55\%& 0.53 & 0.60 & 0.35 \\
			& ROR & 381 && 83 & 12\%& 21\%& 0.66 & 0.26 & 0.23 & & 37 & 18\%& 32\%& 0.82 & 0.39 & \textbf{0.33} & & 16 & 16\%& 36\%& 0.62 & 0.44 & 0.32 & & 0 & 13\%& 39\%& 0.44 & 0.47 & 0.27 \\
			& UOI & 386 && 150 & 17\%& 21\%& 0.54 & 0.23 & 0.22 & & 44 & 36\%& 44\%& 0.88 & 0.49 & \textbf{0.47} & & 17 & 34\%& 45\%& 0.75 & 0.51 & 0.46 & & 1 & 27\%& 51\%& 0.55 & 0.57 & 0.38 \\
			\midrule
			Io & ABS & 251 && 102 & 13\%& 38\%& 0.42 & 0.43 & 0.32 & & 6 & 23\%& 71\%& 0.64 & 0.78 & \textbf{0.53} & & 6 & 23\%& 72\%& 0.58 & 0.79 & 0.50 & & 2 & 21\%& 79\%& 0.50 & 0.83 & 0.43 \\
			& AOR & 387 && 242 & 14\%& 27\%& 0.27 & 0.29 & 0.22 & & 4 & 27\%& 86\%& 0.52 & 0.89 & \textbf{0.51} & & 4 & 27\%& 86\%& 0.49 & 0.90 & 0.49 & & 4 & 27\%& 89\%& 0.40 & 0.92 & 0.38 \\
			& LCR & 446 && 102 & 6\%& 60\%& 0.32 & 0.66 & 0.30 & & 4 & 10\%& 82\%& 0.43 & 0.88 & \textbf{0.43} & & 0 & 10\%& 85\%& 0.39 & 0.90 & 0.41 & & 0 & 9\%& 88\%& 0.30 & 0.92 & 0.32 \\
			& ROR & 454 && 154 & 14\%& 47\%& 0.42 & 0.52 & 0.36 & & 10 & 24\%& 76\%& 0.62 & 0.84 & \textbf{0.57} & & 6 & 24\%& 79\%& 0.55 & 0.87 & 0.54 & & 4 & 20\%& 86\%& 0.41 & 0.91 & 0.41 \\
			& UOI & 351 && 175 & 14\%& 38\%& 0.32 & 0.41 & 0.27 & & 4 & 26\%& 82\%& 0.55 & 0.88 & \textbf{0.53} & & 3 & 26\%& 84\%& 0.51 & 0.90 & 0.51 & & 3 & 25\%& 88\%& 0.41 & 0.92 & 0.41 \\
			\midrule
			Lang & ABS & 278 && 145 & 19\%& 43\%& 0.32 & 0.44 & 0.32 & & 2 & 42\%& 92\%& 0.68 & 0.94 & \textbf{0.70} & & 0 & 42\%& 95\%& 0.64 & 0.98 & \textbf{0.70} & & 0 & 40\%& 96\%& 0.56 & 0.99 & 0.61 \\
			& AOR & 480 && 307 & 9\%& 31\%& 0.20 & 0.32 & 0.19 & & 14 & 40\%& 90\%& 0.68 & 0.91 & 0.66 & & 1 & 40\%& 95\%& 0.61 & 0.99 & \textbf{0.67} & & 0 & 36\%& 95\%& 0.50 & 0.99 & 0.55 \\
			& LCR & 447 && 162 & 16\%& 59\%& 0.33 & 0.60 & 0.35 & & 8 & 27\%& 89\%& 0.55 & 0.90 & 0.55 & & 1 & 27\%& 95\%& 0.48 & 0.97 & \textbf{0.56} & & 0 & 26\%& 95\%& 0.45 & 0.97 & 0.52 \\
			& ROR & 466 && 217 & 22\%& 49\%& 0.34 & 0.51 & 0.37 & & 7 & 43\%& 90\%& 0.67 & 0.94 & \textbf{0.69} & & 1 & 43\%& 94\%& 0.64 & 0.97 & \textbf{0.69} & & 0 & 38\%& 96\%& 0.54 & 0.98 & 0.59 \\
			& UOI & 452 && 242 & 15\%& 40\%& 0.30 & 0.41 & 0.29 & & 8 & 41\%& 88\%& 0.71 & 0.90 & 0.69 & & 0 & 40\%& 93\%& 0.64 & 0.98 & \textbf{0.70} & & 0 & 35\%& 94\%& 0.52 & 0.98 & 0.56 \\
			\midrule
			Gson & ABS & 225 && 19 & 1\%& 38\%& 0.27 & 0.61 & \textbf{0.24} & & 5 & 1\%& 42\%& 0.31 & 0.66 & 0.21 & & 5 & 1\%& 96\%& 0.13 & 0.97 & 0.16 & & 0 & 1\%& 97\%& 0.11 & 0.98 & 0.14 \\
			& AOR & 310 && 1 & 1\%& 57\%& 0.19 & 0.78 & \textbf{0.20} & & 0 & 1\%& 60\%& 0.17 & 0.83 & \textbf{0.20} & & 0 & 1\%& 99\%& 0.09 & 1.00 & 0.13 & & 0 & 0\%& 99\%& 0.07 & 1.00 & 0.12 \\
			& LCR & 431 && 43 & 2\%& 37\%& 0.41 & 0.53 & \textbf{0.22} & & 32 & 2\%& 38\%& 0.47 & 0.56 & 0.21 & & 32 & 1\%& 81\%& 0.24 & 0.88 & 0.19 & & 21 & 0\%& 85\%& 0.20 & 0.89 & 0.15 \\
			& ROR & 703 && 82 & 2\%& 34\%& 0.36 & 0.54 & \textbf{0.22} & & 28 & 2\%& 36\%& 0.43 & 0.58 & 0.21 & & 27 & 2\%& 93\%& 0.17 & 0.94 & 0.17 & & 4 & 2\%& 94\%& 0.15 & 0.95 & 0.15 \\
			& UOI & 418 && 12 & 2\%& 41\%& 0.35 & 0.63 & \textbf{0.24} & & 12 & 2\%& 42\%& 0.34 & 0.68 & \textbf{0.24} & & 12 & 2\%& 92\%& 0.19 & 0.95 & 0.20 & & 9 & 0\%& 94\%& 0.15 & 0.96 & 0.17 \\
			\midrule
			Jgit & ABS & 199 && 90 & 2\%& 16\%& 0.31 & 0.25 & 0.12 & & 1 & 2\%& 35\%& 0.50 & 0.54 & \textbf{0.22} & & 0 & 2\%& 56\%& 0.23 & 0.87 & \textbf{0.22} & & 0 & 2\%& 70\%& 0.11 & 0.94 & 0.12 \\
			& AOR & 386 && 189 & 1\%& 18\%& 0.23 & 0.31 & 0.13 & & 5 & 3\%& 44\%& 0.47 & 0.65 & \textbf{0.27} & & 0 & 2\%& 62\%& 0.23 & 0.90 & 0.24 & & 0 & 2\%& 69\%& 0.14 & 0.94 & 0.15 \\
			& LCR & 334 && 87 & 0\%& 25\%& 0.34 & 0.38 & 0.13 & & 3 & 1\%& 48\%& 0.40 & 0.63 & \textbf{0.20} & & 2 & 1\%& 66\%& 0.19 & 0.88 & \textbf{0.20} & & 0 & 1\%& 76\%& 0.11 & 0.94 & 0.11 \\
			& ROR & 356 && 129 & 2\%& 21\%& 0.34 & 0.33 & 0.15 & & 4 & 3\%& 44\%& 0.49 & 0.62 & \textbf{0.26} & & 1 & 3\%& 62\%& 0.26 & 0.88 & 0.25 & & 0 & 1\%& 73\%& 0.15 & 0.94 & 0.15 \\
			& UOI & 405 && 187 & 1\%& 19\%& 0.25 & 0.28 & 0.11 & & 6 & 3\%& 47\%& 0.45 & 0.64 & \textbf{0.24} & & 2 & 2\%& 65\%& 0.22 & 0.91 & \textbf{0.24} & & 0 & 2\%& 73\%& 0.13 & 0.95 & 0.14 \\
			\midrule
			Jodatime & ABS & 294 && 109 & 14\%& 33\%& 0.47 & 0.35 & 0.25 & & 0 & 19\%& 50\%& 0.74 & 0.56 & 0.39 & & 0 & 19\%& 86\%& 0.38 & 0.97 & \textbf{0.42} & & 0 & 19\%& 89\%& 0.35 & 0.98 & 0.38 \\
			& AOR & 541 && 205 & 11\%& 25\%& 0.50 & 0.27 & 0.19 & & 16 & 13\%& 33\%& 0.82 & 0.38 & 0.27 & & 0 & 11\%& 87\%& 0.28 & 0.97 & \textbf{0.33} & & 0 & 11\%& 89\%& 0.26 & 0.98 & 0.30 \\
			& LCR & 438 && 143 & 8\%& 27\%& 0.46 & 0.33 & 0.18 & & 0 & 9\%& 49\%& 0.61 & 0.57 & 0.30 & & 0 & 9\%& 83\%& 0.29 & 0.95 & \textbf{0.33} & & 0 & 9\%& 84\%& 0.24 & 0.96 & 0.27 \\
			& ROR & 426 && 116 & 16\%& 35\%& 0.57 & 0.39 & 0.28 & & 4 & 19\%& 47\%& 0.76 & 0.54 & 0.38 & & 1 & 19\%& 81\%& 0.41 & 0.93 & \textbf{0.41} & & 0 & 18\%& 85\%& 0.37 & 0.95 & 0.37 \\
			& UOI & 499 && 168 & 13\%& 27\%& 0.53 & 0.30 & 0.22 & & 7 & 15\%& 37\%& 0.80 & 0.42 & 0.30 & & 1 & 13\%& 85\%& 0.31 & 0.97 & \textbf{0.36} & & 1 & 13\%& 86\%& 0.29 & 0.98 & 0.33 \\
			\midrule
			Shindig & ABS & 247 && 51 & 25\%& 54\%& 0.56 & 0.59 & 0.43 & & 19 & 28\%& 67\%& 0.68 & 0.75 & 0.56 & & 19 & 28\%& 81\%& 0.60 & 0.86 & \textbf{0.57} & & 0 & 26\%& 87\%& 0.53 & 0.90 & 0.52 \\
			& AOR & 314 && 61 & 31\%& 57\%& 0.54 & 0.60 & 0.44 & & 26 & 39\%& 75\%& 0.65 & 0.79 & 0.58 & & 22 & 40\%& 87\%& 0.63 & 0.87 & \textbf{0.60} & & 0 & 36\%& 91\%& 0.60 & 0.91 & 0.57 \\
			& LCR & 300 && 25 & 14\%& 54\%& 0.55 & 0.61 & 0.37 & & 25 & 15\%& 60\%& 0.59 & 0.68 & 0.40 & & 12 & 15\%& 71\%& 0.51 & 0.74 & \textbf{0.41} & & 4 & 15\%& 78\%& 0.49 & 0.80 & 0.40 \\
			& ROR & 338 && 36 & 18\%& 48\%& 0.62 & 0.55 & 0.38 & & 23 & 20\%& 54\%& 0.68 & 0.64 & 0.45 & & 18 & 20\%& 68\%& 0.57 & 0.76 & \textbf{0.47} & & 2 & 18\%& 75\%& 0.53 & 0.82 & \textbf{0.47} \\
			& UOI & 251 && 32 & 27\%& 56\%& 0.61 & 0.63 & 0.47 & & 21 & 32\%& 68\%& 0.69 & 0.74 & 0.56 & & 18 & 32\%& 80\%& 0.66 & 0.81 & \textbf{0.57} & & 1 & 30\%& 84\%& 0.62 & 0.86 & 0.55 \\
			\midrule
			Sonarqube & ABS & 288 && 68 & 33\%& 51\%& 0.63 & 0.56 & 0.49 & & 39 & 43\%& 74\%& 0.80 & 0.78 & 0.66 & & 39 & 43\%& 77\%& 0.79 & 0.81 & 0.67 & & 3 & 40\%& 81\%& 0.76 & 0.84 & \textbf{0.68} \\
			& AOR & 253 && 39 & 26\%& 79\%& 0.50 & 0.81 & 0.53 & & 0 & 28\%& 91\%& 0.57 & 0.93 & \textbf{0.59} & & 0 & 28\%& 91\%& 0.57 & 0.93 & \textbf{0.59} & & 0 & 26\%& 91\%& 0.51 & 0.94 & 0.55 \\
			& LCR & 177 && 17 & 25\%& 57\%& 0.70 & 0.63 & 0.50 & & 22 & 27\%& 68\%& 0.75 & 0.73 & \textbf{0.57} & & 22 & 24\%& 76\%& 0.68 & 0.78 & 0.53 & & 5 & 21\%& 78\%& 0.64 & 0.78 & 0.49 \\
			& ROR & 462 && 87 & 30\%& 61\%& 0.60 & 0.65 & 0.51 & & 28 & 34\%& 78\%& 0.70 & 0.82 & 0.62 & & 28 & 34\%& 82\%& 0.69 & 0.85 & \textbf{0.63} & & 7 & 33\%& 84\%& 0.67 & 0.86 & 0.61 \\
			& UOI & 172 && 20 & 28\%& 72\%& 0.59 & 0.75 & 0.51 & & 9 & 31\%& 81\%& 0.66 & 0.85 & \textbf{0.58} & & 9 & 30\%& 84\%& 0.64 & 0.86 & \textbf{0.58} & & 1 & 28\%& 85\%& 0.59 & 0.87 & 0.54 \\
			\midrule
			Spojo & ABS & 8 && 0 & 0\%& 62\%& 0.42 & 0.62 & \textbf{0.08} & & 0 & 0\%& 62\%& 0.42 & 0.62 & \textbf{0.08} & & 0 & 0\%& 62\%& 0.42 & 0.62 & \textbf{0.08} & & 0 & 0\%& 62\%& 0.42 & 0.62 & \textbf{0.08} \\
			& AOR & 20 && 0 & 0\%& 0\%& 1.00 & 0.00 & 0.00 & & 0 & 0\%& 10\%& 0.45 & 0.13 & \textbf{0.12} & & 0 & 0\%& 10\%& 0.45 & 0.13 & \textbf{0.12} & & 0 & 0\%& 10\%& 0.45 & 0.13 & \textbf{0.12} \\
			& LCR & 48 && 0 & 0\%& 92\%& 0.31 & 0.96 & \textbf{0.41} & & 0 & 0\%& 92\%& 0.31 & 0.96 & \textbf{0.41} & & 0 & 0\%& 100\%& 0.28 & 1.00 & 0.40 & & 0 & 0\%& 100\%& 0.27 & 1.00 & 0.38 \\
			& ROR & 142 && 0 & 2\%& 67\%& 0.55 & 0.69 & 0.36 & & 0 & 5\%& 72\%& 0.46 & 0.78 & \textbf{0.45} & & 0 & 4\%& 86\%& 0.36 & 0.90 & 0.44 & & 0 & 4\%& 90\%& 0.31 & 0.92 & 0.39 \\
			& UOI & 15 && 0 & 0\%& 73\%& 0.63 & 0.76 & \textbf{0.49} & & 0 & 0\%& 73\%& 0.63 & 0.76 & \textbf{0.49} & & 0 & 0\%& 80\%& 0.59 & 0.80 & \textbf{0.49} & & 0 & 0\%& 80\%& 0.58 & 0.80 & 0.48 \\
			\bottomrule
		\end{tabular}
	\end{table}
\end{landscape}
}

A high recall value indicates that the prediction includes the actual impacted test cases. Thus, the complete set value is strongly linked with the recall. 
Indeed, we observe that the complete set value ($p_\C$) is high when the recall value ($R$) is high. 
This is also a piece of evidence of the correctness of our experimental evaluation technique.

	\cg gives the best precision values of all other call graphs. Precision decreases when taking into account the hierarchy or the access to fields (\cgcha and \cgf). This makes us think that more nodes/edges are added, more impacted test cases can be found, which also means, more false positives.
	
	Moreover, the precision varies greatly depending on the mutation operators and project: if we consider \cgcha, it goes from lower values such as 0.09 for Gson with AOR mutation operator to higher ones such as 0.79 for Sonarqube with ABS mutation operator. This observation underlines again the project-dependent side of the impact prediction technique  (see Research~Question~\ref{rq:projdep}).

\takeaway{
	The four types of call graph under consideration are not equivalent for impact prediction.
	According to our protocol, the best one is \cg, which does not consider Class Hierarchy Analysis and field analysis.
	The main reason is that the sophistication of Class Hierarchy Analysis and field analysis increases the recall of impact prediction but decreases too much the precision.
}

\RQ{\rqprojmutdep}
\label{rq:projdep}

Let us again consider Table \ref{table:metrics_for_graphs}. Now, we focus on the difference between projects and mutation operators: 
\begin{inparaenum}[(i)]
	\item the values differ strongly from one project to another for a given mutation operator (\eg considering the ABS mutation operator with \cg, 3\% in $p_\S$ for Apache Commons Codec, 19\% for Jodatime and 43\% for Sonarqube);
	\item the values differ less from a mutation operator to another for a given project (\eg{considering the Apache Commons Codec  with \cg, values range from 2\% in $p_\S$ for LCR mutation operator to 5\% for ROR mutation operator}).
\end{inparaenum}

These observations highlight the fact that the accuracy of the call graph impact prediction technique depends more on the project than on the mutation operator. Though instantiated through software projects, this observation really concerns the architecture of the software project, or the development patterns (\eg extensive usage of hierarchy, of reflection, \etc) employed to realize the project.

Similar observations have already been reported in Research~Question~\ref{rq:perf}: the fact that a call graph shows better F-scores for all mutations operators of the project and the fact that the precision may have really low or high values depending on the project.

\takeaway{Call graph-based impact prediction is influenced by the structure of call graphs and by the mutation operators used in the experiment. The project-dependence is higher than the mutation operator dependence.}

\RQ{\rqreasonbadaccuracy}

The answer to Research Question~\ref{rq:perf} has reported both low and high accuracy of impact prediction using call graphs.
To gain even better knowledge about call graphs, we have performed an investigation on a set of cases for which the prediction error is particularly bad. We now discuss our main findings.

Our technique is based on a static call graph. Hence, the call graph does not handle the use of Java Reflection mechanism. The reflection mechanism is resolved at run time while the call graph is built in a static manner. Obviously, this leads to discrepancies between the results of our analysis, and the outcome of the execution.
However, we may detect the use of reflection in a project since then, the source code refers to some specific classes/packages in the Java library (package \texttt{java.reflect}). In practice, we may raise a warning to the user about the use of reflection mechanism so that he would take special care when interpreting the impact.

We also notice that the test cases are not independent from each others.
Since mutation analysis is costly, we execute them in parallel.
In the case of Apache Common Io, the parallel execution of tests sometimes results in failing test cases, where the failure is due to parallel execution and not the mutation itself.
The reason is that Apache Common Io extensively uses the hard drive. As our parallel  test cases run on the same hard drive space (\ie folder), they try to read/write/create identical folders/files. Consequently, some test cases fail due to this parallel I/O but it is not due to the mutant itself. There are different ways to address this problem: the easiest one is to run one instance of test at a time in a manner that the I/O is not shared. Another way is to duplicate the project for each mutation operator in a way that if they run in parallel, each one benefits of its own drive space.

\takeaway{The bad accuracy is related to a low recall and/or a low precision. The low recall of call graph-based impact prediction is caused by missing edges in the graphs (\eg because of reflection). The low precision is caused by too many edges in the considered graphs, especially for \cgcha and \cgf.}

\RQ{\rqtime}
\label{rq:times}

Now that we have a clearer understanding of the precision and recall of call graph-based impact prediction, we concentrate on the execution time of the prediction.

Table~\ref{table:times} gives the computation time for each project (column 1) of our dataset. Each time related to the call graph is given for the four types of call graph (\cgj, \cg, \cgcha and \cgf).
These times are:
\begin{inparaenum}[(i)]
    \item $t_{test}$, the time required  to run the test suite (column 2);
    \item the time required to build the call graph for each call graph type (columns 3, 5, 7, and 9);
    \item the average time of impact prediction based on the call graph, \ie computing one impact prediction (column 4, 6, 8, 10).
\end{inparaenum}
The average time of impact prediction is expressed in milliseconds, for instance it takes 0.11 millisecond in average to make an impact prediction in Apache Commons Codec with \cgcha.

\begin{table}
  \caption{Main computation time to run each test suite, to build each call graph and to predict one impact using each call graph.}
  \label{table:times}
  \centering
  \begin{tabular}{@{}lcrcrrcrrcrrcrr@{}}
    \toprule
    &&&& \multicolumn{2}{c}{\cgj} && \multicolumn{2}{c}{\cg} && \multicolumn{2}{c}{\cgcha} && \multicolumn{2}{c}{\cgf} \\
    \cline{5-6}
    \cline{8-9}
    \cline{11-12}
    \cline{14-15}
    Project && $t_{test}$ && build & pred. && build & pred. && build & pred. && build & pred. \\
    \midrule
    Codec && 32.2s && 3h+ & 0.09ms && 0.78s & 0.04ms && 0.96s & 0.11ms && 0.90s & 0.17ms \\
    Collections && 38.9s && 9h+ & 2.03ms && 4.14s & 0.03ms && 3.78s & 0.08ms && 3.98s & 0.48ms \\
    Gson && 13.7s && 2h+ & 0.34ms && 1.01s & 0.54ms && 1.16s & 1.66ms && 0.90s & 3.23ms  \\
    Io && 90.1s && 3h+ & 0.21ms && 1.51s & 0.05ms && 0.91s & 0.05ms && 0.86s & 0.35ms \\
    Jgit && 195.5s &&  40h+ & 5.25ms && 10.80s & 0.99ms && 6.52s & 3.48ms && 6.03s & 40.29ms \\
    Jodatime && 31.2s && 25h+ & 2.55ms && 8.12s & 0.61ms && 4.92s & 10.50ms && 4.36s & 23.14ms \\
    Lang && 40.0s && 15h+ & 1.81ms && 2.82s & 0.05ms && 2.75s & 0.06ms && 2.75s & 0.31ms \\
    Shindig && 14.1s && 1h+ & 0.17ms && 0.65s & 0.02ms && 0.58s & 0.04ms && 0.63s & 0.08ms \\
    Sonarqube && 387.3s && 3h+ & 0.76ms && 1.74s & 0.02ms && 1.42s & 0.01ms && 1.25s & 0.10ms \\
    Spojo && 2.7s && 16m & 0.06ms && 0.22s & 0.04ms && 0.24s & 0.07ms && 0.25s & 0.12ms \\
    \bottomrule
  \end{tabular}
\end{table}

First, we observe the time needed to generate call graphs with JavaPDG (\cgj). We observe that it takes several hours to generate the call graph with all elements.
For the smallest project, Spojo, it takes 16 minutes. 
For the largest, Jgit, it takes almost 2 days of computation. 
This aspect is linked to the fact that JavaPDG also builds a finer graph (the Program Dependence Graph) before extracting the call graph.
Thus, using JavaPDG has an important cost in time.

If we focus on \cg, \cgcha and \cgf, we observe that building these graphs takes from 1 to 11 seconds (with an average time of 2.6 seconds) for all projects and call graphs being considered. Furthermore, it takes less than 5 seconds for almost all projects (except for Jgit and Jodatime which are the two largest projects, which both require respectively up to 10.8s and 8.1s). 
The building process seems to last longer with lighter types of the graph (\ie \cg) than with heavier ones (\ie \cgf).
However, our implementation always lists all elements, but just filter out some nodes depending on the type of graph. 
Thus, these differences in time are more probably explained by the system load at the moment of the generation.

Once the graph is built, determining an impact takes less than 45ms for all projects, and even less than 5ms for all projects except again for Jgit and Jodatime. 
These observations also apply to graphs generated using JavaPDG: all predictions are made in less than 5ms (except for Jgit: 5.25ms).
We observe that prediction times using JavaPDG are generally larger. 
These differences are likely to be related to the fact that the graph is obtained by third-party software, the data returned is not exactly the same as ours.
Thus, some additional on-the-fly data transformations are required to find good nodes in the graph.
Overall, these prediction times are equivalent.

If we compare the prediction for \cg, \cgcha and \cgf, we can see that the prediction time increases with the size of the software and the graph. Thus, \cgf which is the largest graph (as it contains more nodes and edges) increases the required time for prediction, but this increase remains reasonable (maximum absolute value of 41ms). This is expected since prediction is based on path enumeration in the graph.
Prediction with a lighter version of the call graph (\cg) performs impact analysis in less than 1ms for all cases.

One can determine actual impacts directly by running the program test cases.
Thus, if we look at the time required for the execution of the test cases, we see that the minimal time required is 3 seconds for the smallest project (Spojo) and can reach values as high as 387.3 seconds for Sonarqube. 
The time required to build the call graph is smaller to the time required to run test cases for software. 
This shows that using call graphs to predict impacts costs less than running test suites. If we consider Apache Common Codec, the time required to build a call graph is more than 33 times smaller than the time required to run the entire test suite.
And by comparison, the time to make a prediction is orders of magnitude smaller than the time to run the test suite.
This observation is interesting as it underlines the fact that using an impact prediction technique based on a call graph can quickly provide some insight of the consequences of a change on the tests. One can first run the returned test to directly find failing ones, which represents a gain of time for the developer.

Furthermore, the call graph has the advantage that during software evolution, many changes would have no impact on it (\eg changing an operator, shifting a line in the code, \etc). 
Thus, the same call graph can be used for predicting the impact of several simultaneous changes before requiring graph regeneration. 
Moreover, when it is required to recompute it, the time to build the call graph is reasonable, within seconds for our dataset with a maximum of 11 seconds for Jgit.	
This makes it possible to use such an impact prediction on the fly in the development environment (IDE).

This opens interesting research avenues, where one first performs very fast approximation of error propagation before performing more sophisticated static analyses. This can even be used in a pre-processing step for a dynamic analysis. 

Now consider that large companies have hundreds of thousands of interrelated test cases, as in the case of Google \cite{seo_programmers_2014}.
It is likely that these scenarios will be more and more common, and that low-level, detailed analysis of the computation will fail to scale.
We think that such settings will need very fast approximation of impacts. 
The preliminary performance results we report here, with un-optimized software, make us confident that this is indeed possible.

\takeaway{
	Call graph-based impact prediction is orders of magnitude faster than actually running the test suite. The time cost to build the call graph is also much smaller.
	In a software codebase with a very large number of methods and test cases, the imprecision of call graph-based impact prediction is compensated by the gain in execution time.
	Passing from \cg to \cgf makes prediction times slower, but these times remains acceptable for prediction and much faster than running test cases.}

\subsection{Discussion}

\subsubsection{Other Software Graphs}
\label{sec:othercg}

In this section, we motivate our choice of the call graph as a change impact analysis technique. 
Other software graphs can be used for impact prediction. 
Two examples discussed here are:
\begin{inparaenum}[(i)]
    \item the program dependence graph (\aka PDG) which contains more nodes/edges than the call graph as it contains more low-level elements (\ie code instructions) from the source code;
    \item the class or the package dependency graph which contains less nodes/edges than the call graph as it contains only dependencies between classes or packages. That is, an edge is added between a class or a package \texttt{A} and a class or a package \texttt{B} every time a method of \texttt{A} access to any element (\ie class, method, field, constant, \etc) of the class or package \texttt{B}.
\end{inparaenum}

If we consider a finer granularity graph (such as the program dependence graph, \aka PDG), it will hardly scale with large programs. 
This intuition is validated with our first experiments: the time required for building PDG with the well-known program JavaPDG are important (\cf Table~\ref{table:times}). 
Building a graph for all the programs of our experimental dataset took more than 4 days (with an average time per project of more than 10 hours).

To the opposite, coarser granularity (such as a class or package dependence graph) contains less information.
An impact prediction for a change in a method \texttt{Pkg.Foo.bar()} is interpreted as a change introduced in the \texttt{Pkg.Foo} class or in the package \texttt{Pkg}.
As a consequence, the resulting impacts are inevitably of the same granularity (\ie classes or packages).
This results in considering all tests of a test class (or a package) as also failing.
That is, considering the amount of methods or fields a class can contain (and the number of class a package can contain), the resulting prediction will be inevitably bad,  and it would be difficult to precisely locate the impacts.
To better understand this point, we have computed the class and package dependency graphs for the projects in our dataset. We observe that we have more or less 10 times fewer nodes in a dependency graph than in a call graph, and less than 30 nodes in the package dependency graph. Similar observations can be made regarding the edges.
Now if we consider the smallest project (Spojo) which contains 330 methods (nodes) and 890 calls (edges), we observe that the class dependency graph contains only 37 classes (nodes) and 69 dependencies (edges). These figures get even worse with the package dependency graph which contains only 7 packages (nodes) and 13 dependencies (edges).

To sum up, we use call graphs for impact prediction because it exhibits a good trade-off between performance and cost.
Moreover as a test is a method, it is also a natural unit of decomposition.

\subsubsection{Comparison against Impact Prediction Techniques}

In this paper, we focus on characterizing the efficiency of different call graphs for impact prediction (depending on which features we include in the call graph computation -- inheritance and fields).
Comparing the accuracy of this technique to existing ones is another research question.
We wanted to answer to such a research question but this is impossible so far.
We identify two reasons that make such a comparative study a challenge.

The first reason is that the proposed tools do not necessarily work at the same granularity and/or language. As an example, some may observe code statements of C language \cite{ramanathan_sieve:_2006}.
The second reason is that the techniques which can be compared to ours \cite{li_survey_2013} do not provide a publicly available implementation (even by contacting directly the authors). 
The latter reason is why we make all our implementation publicly available.
To sum up, due to the lack of open tools, a comparative evaluation of impact prediction on Java software at the level of methods is not possible.

\subsubsection{Threats to Validity}

At a conceptual level, the main threat to the validity of our experimental results is that we consider the test suite execution as ground truth.  However, it may be the case that the test cases miss the assertions that would detect the actually propagated error and thus fail. This threat is mitigated by our manual analysis.

Our large scale experiment confirms known and yet essential facts to be taken into account when doing mutation analysis. One of such a consideration is the fact a single mutant sometimes makes an entire test suite broken. 
As an example, if one uses a static field in a test class which is initialized by default with a mutated constructor then, if the mutation has made the constructor ineffective, it results in an unexpected behavior and an entire test class cannot be initialized. In such a situation, the test suite is reported as failing, and consequently, all test cases belonging to the test suite are reported similarly.
	
Another example is the fact that a test may hang. Indeed, let us imagine the mutation changes a loop condition which results in an infinite looping. To circumvent this problem, we add a timeout for each test. This way, we can determine if some hangs or not. It is equally important to use a reasonable timeout value for the project to avoid considering a test as hanging when it is not.

\subsection{Qualitative Comparison}
\label{sec:qualitativecomp}

In this section, we discuss the most closely related work.
Law and Rothermel \cite{law_whole_2003} have proposed an approach for impact analysis; their technique is based on a code instrumentation to analyze execution stack traces. They compare their technique against simple call graphs on a small piece of software. Their evaluation is based on faults for only one project, and our experiment is much larger. Our technique is evaluated on \nbprojects different software packages.

Hattori \etal \cite{hattori_precision_2008} have used an approach based on call graphs to study propagation. Their evaluation is made on a small dataset made of three projects. Their goal was to show that precision and recall are good tools as evaluation of the performance for an impact analysis technique. We build on their work our evaluation metrics. Our key novelty is that we propose to use mutants for evaluation, and our study has much more subjects: 10 large scale open-source projects.

Cai \etal \cite{cai_sensa:_2014} have proposed a novel technique for impact prediction. Compared to ours, their technique is dynamic and it requires a costly instrumentation phase. In our work, the motto is to have a very fast technique without instrumenting the code, which gives a good approximation as shown by our experimental results. Their implementation is not publicly available for a quantitative comparison.

\section{Related Work}
\label{sec:relworks}

Mutation testing is an old concept which has seen many contributions over years. Jia and Harman propose a survey regarding this topic \cite{jia_analysis_2011}. In this section, we focus on the work that is related to ours. The most related work has already been discussed in \ref{sec:qualitativecomp}.

Strug and Strug \cite{strug_machine_2012} use control flow graphs and classification for detecting similar mutants. Their approach is intended to reduce the number of mutants considered when doing mutation testing. We use these tools for change impact analysis.

Do and Rothermel \cite{do_controlled_2005} describe a protocol to study test case prioritization techniques based on mutation.
Their protocol and ours share the same idea, that of using test cases to determine which test cases are impacted by the change. However, we have a different goal: they study test case prioritization whereas we study impact prediction.

Change impact analysis has been studied for many years and many algorithms have been proposed. 
Many categorizations of such algorithms exist. 
Bohner and Arnold proposed two types of analysis: dependency analysis and traceability analysis \cite{bohner_software_1996}. 
The former analyzes the source code of the program at a relatively fine granularity (\eg methods call, data usage, control statements, \dots) while the latter compares elements at a coarser granularity such as documentation and specifications (\eg UML, \etc). 
Moreover, different types of impact determination techniques are presented. 
According to this paper, our approach is a dependency analysis based on a transitive closure technique.

Bohner and Arnold \cite{bohner_software_1996} and Li \etal \cite{li_survey_2013} list the notable graph-based approaches.
Different types or variants of software graphs have been used to perform change impact analysis, a common example is the program dependence graphs (\aka PDG) \cite{loyall_using_1993}. In the present paper, we focus on the call graph.

Walker \etal \cite{walker_lightweight_2006} propose an impact analysis tool named TRE. Their approach uses conditional probability dependency graphs in which a node represents a class, there is an edge from a node/class A to node B if A contains anything resolving to B. The conditional probabilities are estimated from data extracted from the CVS repository; more precisely, these conditional probabilities are estimated by the number of times two classes are changed on the same commit. Then, the impact of a change is determined based on the resulting graphical model. They work at the level of classes and give no concrete information about the evaluation. In contrast to this work, we work at a finer granularity (methods) which gives us more realistic data and we report numerical evidence for \nbprojects Java packages.

Zimmermann and Nagappan \cite{zimmermann_predicting_2008} propose to use dependency graphs to estimate the most critical parts of a piece of software. Their approach uses network measures and complexity metrics to make the predictions. They assess their findings using some popular though proprietary software, where they are able to determine parts of the software that can cause issues. In contrast, we propose a technique to determine which parts of a piece of software will be impacted by a potential change. Moreover, we experiment our approach on \nbprojects different open-source software packages.

Antoniol \etal \cite{antoniol_identifying_2000} also address impact analysis. However, they consider a slightly different problem setting, because they take as input a bug report or a modification request and not a single source code element as we do. Their approach is less accurate as it takes into consideration documentation (\ie bug reports) for change impact analysis. Our approach is more realistic as it is source-code centric: we only deal with existing elements obtained from source code. The same argument applies for the recent work by Gethers and colleagues \cite{gethers_integrated_2012}.

A classical paper by Moritoni and Winkler \cite{moriconi_approximate_1990} also studies error propagation but they do it with the goal of having a perfect aproximation.
By contrast, we perform approximations with the goal of exploring other trade-offs between precision and recall for impact prediction. Their work is more theoretical in essence, only on small toy examples, whereas we propose a study on real large-scale open-source source code.

Michael and Jones \cite{michael_uniformity_1997} alter variables during the program's execution in order to study how this affects (``perturbates'' in their phrasing) the software. They focus on data-state perturbation, where we have a more global look of the software. Considering only variable perturbation does not take into consideration all the ways an error can propagate. According to our experiments, call egdes better reflect propagation than variable edges.

Challet and Lombardoni \cite{challet_bug_2004} propose a theoretical reflection about impact analysis using graphs. However, they do not evaluate the validity of their ``bug basins'' as we do in this paper.

Robillard and Murphy \cite{robillard_concern_2002} introduce ``concern graphs'' for reasoning on the implementation of features. This kind of graphs may be assessed with the protocol we have presented here.

Binkley \etal \cite{binkley_orbs:_2014,binkley_orbs_2015} propose observation-based slicing (\aka ORBS). They propose to slice a piece of software in a ``delete--execute--observe'' paradigm. In this paradigm, the effects of a change are observed after executing the code (\eg by running test cases). This paradigm is comparable to our approach where we mutate and then run tests to observe the impacts. However, these two techniques are totally different from each other: their technique focuses on a quite low granularity (statements) which makes their approach resource demanding. Our approach has the advantage to be light enough so that one can use it to do run time prediction.

Ren \etal \cite{ren_chianti:_2004} propose a tool entitled Chianti for change impact prediction as an Eclipse plug-in. 
However, beyond the common idea of reasoning about impacts, they target a completely different problem: we aim at finding sensitive methods, while they aim at finding the change responsible for a failure (\aka a bug-inducing commit). Naturally, their techniques and evaluation follow completely different paths.

\section{Conclusion}
\label{sec:conclusion}

Predicting the impact of a change in a piece of software is an important matter.
In this paper, we use different types of call graphs to predict the software elements that are likely to be impacted by a change in the software.
For that purpose, we have introduced new variants of call graph.
The goal of these new call graphs is to perform graph-based impact prediction.
The different types of call graph we use contain an increasing amount of information, being more and more precise in their modeling of the interactions of the elements of a piece of software.
We predict the impact of software changes by navigating in the graph, from the source of change to the different elements that may be reached.
More information in the call graph leads to better prediction; however, once the computation effort comes into consideration, the trade-off between the computation cost and the accuracy of the prediction leads to interesting insights. Then, we are able to discuss whether one favors a fast approximation of the software elements to which a software change may propagate, or a slow, more precise, such prediction.
To discuss these issues based on solid and practical grounds, we present a protocol for experimentally assessing the accuracy of call graphs for impact analysis.
This novel technique, based on mutation testing, is fully automated and enables us to compute standard precision and recall measures.
Specifically, we have executed our protocol on \nbprojects mainstream open-source Java software packages.
The analysis of the predicted impact of \nbmutants mutants shows that one of the call graphs provides a good trade-off between precision and recall. Moreover, this call graph offers good execution times; this let us use it in real execution scenarios such as real time tools for assisting a developer while he is editing his source code; it may also be used as a tool for regression test selection.

Whether based on a static source code analysis, or the analysis of dynamic traces of execution, impact prediction is bound to imperfection. Performing a static analysis, some interactions between software elements cannot be correctly estimated (unless one assumes that any software can interact with any other one, which would obviously lead to far too many predicted impacts, that is far too many false positives). Hence, we do not have all information we need to be perfectly accurate in our estimation of the impacts. Situations where the lack of information has to be dealt with is an essential feature of machine learning; another essential feature of machine learning is to design approximate procedures to estimate quantities of interest, using reasonable computational resources, while keeping guarantees on the overall quality and soundness of the approach. Henceforth, our future work involves applying machine learning techniques to learn the paths in the call graphs where the errors propagate. This goal may be reached by using more information to characterize software elements (nodes) and their interactions (edges), and take advantage of this information to improve the estimation of the paths that propagate different types of software changes.

\bibliographystyle{spmpsci}
\bibliography{biblio}

\begin{thebibliography}{10}
\providecommand{\url}[1]{{#1}}
\providecommand{\urlprefix}{URL }
\expandafter\ifx\csname urlstyle\endcsname\relax
  \providecommand{\doi}[1]{DOI~\discretionary{}{}{}#1}\else
  \providecommand{\doi}{DOI~\discretionary{}{}{}\begingroup
  \urlstyle{rm}\Url}\fi

\bibitem{acharya_practical_2012}
Acharya, M., Robinson, B.: Practical {Change} {Impact} {Analysis} {Based} on
  {Static} {Program} {Slicing} for {Industrial} {Software} {Systems}.
\newblock In: Proceedings of the 20th {International} {Symposium} on the
  {Foundations} of {Software} {Engineering}, {FSE} '12, pp. 13:1--13:2. ACM,
  New York, NY, USA (2012).
\newblock \doi{10.1145/2393596.2393610}

\bibitem{antoniol_identifying_2000}
Antoniol, G., Canfora, G., Casazza, G., de~Lucia, A.: Identifying the
  {Starting} {Impact} {Set} of a {Maintenance} {Request}: {A} {Case} {Study}.
\newblock In: Proceedings of the {Conference} on {Software} {Maintenance} and
  {Reengineering}, {CSMR} '00, pp. 227--. IEEE Computer Society, Washington,
  DC, USA (2000)

\bibitem{arnold_impact_1993}
Arnold, R.S., Bohner, S.A.: Impact {Analysis} - {Towards} a {Framework} for
  {Comparison}.
\newblock In: Proceedings of the {Conference} on {Software} {Maintenance},
  {ICSM} '93, pp. 292--301. IEEE Computer Society, Washington, DC, USA (1993)

\bibitem{binkley_orbs:_2014}
Binkley, D., Gold, N., Harman, M., Islam, S., Krinke, J., Yoo, S.: {ORBS}:
  {Language}-independent {Program} {Slicing}.
\newblock In: Proceedings of the 22Nd {ACM} {SIGSOFT} {International}
  {Symposium} on {Foundations} of {Software} {Engineering}, {FSE} 2014, pp.
  109--120. ACM, New York, NY, USA (2014).
\newblock \doi{10.1145/2635868.2635893}

\bibitem{binkley_orbs_2015}
Binkley, D., Gold, N., Harman, M., Islam, S., Krinke, J., Yoo, S.: {ORBS} and
  the {Limits} of {Static} {Slicing}.
\newblock In: 2015 {IEEE} 15th {International} {Working} {Conference} on
  {Source} {Code} {Analysis} and {Manipulation} ({SCAM}), pp. 1--10 (2015).
\newblock \doi{10.1109/SCAM.2015.7335396}

\bibitem{bohner_software_2002}
Bohner, S.: Software {Change} {Impacts} - {An} {Evolving} {Perspective}.
\newblock In: Proceedings of the {International} {Conference} on {Software}
  {Maintenance}, {ICSM} '02, pp. 263--272 (2002).
\newblock \doi{10.1109/ICSM.2002.1167777}

\bibitem{bohner_software_1996}
Bohner, S.A., Arnold, R.S.: Software {Change} {Impact} {Analysis}.
\newblock IEEE Computer Society Press, Los Alamitos, CA, USA (1996)

\bibitem{cai_sensa:_2014}
Cai, H., Jiang, S., Santelices, R., Zhang, Y.J., Zhang, Y.: {SENSA}:
  {Sensitivity} {Analysis} for {Quantitative} {Change}-{Impact} {Prediction}.
\newblock In: Proceedings of the 14th {International} {Working} {Conference} on
  {Source} {Code} {Analysis} and {Manipulation}, {SCAM} '14, pp. 165--174. IEEE
  Computer Society, Washington, DC, USA (2014).
\newblock \doi{10.1109/SCAM.2014.25}

\bibitem{challet_bug_2004}
Challet, D., Lombardoni, A.: Bug {Propagation} and {Debugging} in {Asymmetric}
  {Software} {Structures}.
\newblock Physical Review E \textbf{70}(4), 046,109 (2004).
\newblock \doi{10.1103/PhysRevE.70.046109}

\bibitem{dean_optimization_1995}
Dean, J., Grove, D., Chambers, C.: Optimization of {Object}-{Oriented}
  {Programs} {Using} {Static} {Class} {Hierarchy} {Analysis}.
\newblock In: Proceedings of the 9th {European} {Conference} on
  {Object}-{Oriented} {Programming}, {ECOOP} '95, pp. 77--101. Springer-Verlag,
  London, UK, UK (1995)

\bibitem{do_controlled_2005}
Do, H., Rothermel, G.: A {Controlled} {Experiment} {Assessing} {Test} {Case}
  {Prioritization} {Techniques} via {Mutation} {Faults}.
\newblock In: Proceedings of the 21st {International} {Conference} on
  {Software} {Maintenance}, {ICSM} '05, pp. 411--420. IEEE Computer Society,
  Washington, DC, USA (2005).
\newblock \doi{10.1109/ICSM.2005.9}

\bibitem{gethers_integrated_2012}
Gethers, M., Dit, B., Kagdi, H., Poshyvanyk, D.: Integrated {Impact} {Analysis}
  for {Managing} {Software} {Changes}.
\newblock In: Proceedings of the 34th {International} {Conference} on
  {Software} {Engineering}, {ICSE} '12, pp. 430--440. IEEE Press, Piscataway,
  NJ, USA (2012)

\bibitem{grove_call_1997}
Grove, D., DeFouw, G., Dean, J., Chambers, C.: Call {Graph} {Construction} in
  {Object}-oriented {Languages}.
\newblock In: Proceedings of the {Conference} on {Object}-oriented
  {Programming}, {Systems}, {Languages}, and {Applications}, pp. 108--124
  (1997)

\bibitem{hattori_precision_2008}
Hattori, L., Guerrero, D., Figueiredo, J., Brunet, J., Damásio, J.: On the
  {Precision} and {Accuracy} of {Impact} {Analysis} {Techniques}.
\newblock In: Proceedings of the {Seventh} {IEEE}/{ACIS} {International}
  {Conference} on {Computer} and {Information} {Science} ({Icis} 2008), {ICIS}
  '08, pp. 513--518. IEEE Computer Society, Washington, DC, USA (2008).
\newblock \doi{10.1109/ICIS.2008.104}

\bibitem{jia_analysis_2011}
Jia, Y., Harman, M.: An {Analysis} and {Survey} of the {Development} of
  {Mutation} {Testing}.
\newblock IEEE Transactions on Software Engineering \textbf{37}(5), 649--678
  (2011).
\newblock \doi{10.1109/TSE.2010.62}

\bibitem{king_fortran_1991}
King, K.N., Offutt, A.J.: A {Fortran} {Language} {System} for {Mutation}-based
  {Software} {Testing}.
\newblock Software: Practice and Experience \textbf{21}(7), 685--718 (1991).
\newblock \doi{10.1002/spe.4380210704}

\bibitem{law_whole_2003}
Law, J., Rothermel, G.: Whole {Program} {Path}-{Based} {Dynamic} {Impact}
  {Analysis}.
\newblock In: Proceedings of the 25th {International} {Conference} on
  {Software} {Engineering}, {ICSE} '03, pp. 308--318. IEEE Computer Society,
  Washington, DC, USA (2003)

\bibitem{lehnert_taxonomy_2011}
Lehnert, S.: A {Taxonomy} for {Software} {Change} {Impact} {Analysis}.
\newblock In: Proceedings of the 12th {International} {Workshop} on
  {Principles} of {Software} {Evolution} and the 7th {Annual} {ERCIM}
  {Workshop} on {Software} {Evolution}, {IWPSE}-{EVOL} '11, pp. 41--50. ACM,
  New York, NY, USA (2011).
\newblock \doi{10.1145/2024445.2024454}

\bibitem{li_survey_2013}
Li, B., Sun, X., Leung, H., Zhang, S.: A {Survey} of {Code}-based {Change}
  {Impact} {Analysis} {Techniques}.
\newblock Software Testing, Verification and Reliability \textbf{23}(8),
  613--646 (2013).
\newblock \doi{10.1002/stvr.1475}

\bibitem{loyall_using_1993}
Loyall, J.P., Mathisen, S.A.: Using {Dependence} {Analysis} to {Support} the
  {Software} {Maintenance} {Process}.
\newblock In: Proceedings of the {Conference} on {Software} {Maintenance},
  {ICSM} '93, pp. 282--291. IEEE Computer Society, Washington, DC, USA (1993)

\bibitem{michael_uniformity_1997}
Michael, C.C., Jones, R.C.: On the {Uniformity} of {Error} {Propagation} in
  {Software}.
\newblock In: Proceedings of the 12th {Annual} {Conference} on {Computer}
  {Assurance}, {COMPASS}'97, pp. 68--76 (1997).
\newblock \doi{10.1109/CMPASS.1997.613237}

\bibitem{moriconi_approximate_1990}
Moriconi, M., Winkler, T.C.: Approximate {Reasoning} {About} the {Semantic}
  {Effects} of {Program} {Changes}.
\newblock IEEE Transactions on Software Engineering \textbf{16}(9), 980--992
  (1990).
\newblock \doi{10.1109/32.58785}

\bibitem{offutt_experimental_1996}
Offutt, A.J., Lee, A., Rothermel, G., Untch, R.H., Zapf, C.: An {Experimental}
  {Determination} of {Sufficient} {Mutant} {Operators}.
\newblock ACM Transactions on Software Engineering and Methodology
  \textbf{5}(2), 99--118 (1996).
\newblock \doi{10.1145/227607.227610}

\bibitem{pawlak_spoon:_2015}
Pawlak, R., Monperrus, M., Petitprez, N., Noguera, C., Seinturier, L.: Spoon:
  {A} {Library} for {Implementing} {Analyses} and {Transformations} of {Java}
  {Source} {Code}.
\newblock Software: Practice and Experience p.~na (2015).
\newblock \doi{10.1002/spe.2346}

\bibitem{ramanathan_sieve:_2006}
Ramanathan, M.K., Grama, A., Jagannathan, S.: Sieve: {A} {Tool} for
  {Automatically} {Detecting} {Variations} {Across} {Program} {Versions}.
\newblock In: Proceedings of the 21st {IEEE}/{ACM} {International} {Conference}
  on {Automated} {Software} {Engineering}, {ASE} '06, pp. 241--252. IEEE
  Computer Society, Washington, DC, USA (2006).
\newblock \doi{10.1109/ASE.2006.61}

\bibitem{ren_chianti:_2004}
Ren, X., Shah, F., Tip, F., Ryder, B.G., Chesley, O.: Chianti: {A} {Tool} for
  {Change} {Impact} {Analysis} of {Java} {Programs}.
\newblock In: Proceedings of the 19th {Annual} {ACM} {SIGPLAN} {Conference} on
  {Object}-oriented {Programming}, {Systems}, {Languages}, and {Applications},
  {OOPSLA} '04, pp. 432--448. ACM, New York, NY, USA (2004).
\newblock \doi{10.1145/1028976.1029012}

\bibitem{robillard_concern_2002}
Robillard, M.P., Murphy, G.C.: Concern {Graphs}: {Finding} and {Describing}
  {Concerns} {Using} {Structural} {Program} {Dependencies}.
\newblock In: Proceedings of the 24th {International} {Conference} on
  {Software} {Engineering}, {ICSE} '02, pp. 406--416. ACM, New York, NY, USA
  (2002).
\newblock \doi{10.1145/581339.581390}

\bibitem{seo_programmers_2014}
Seo, H., Sadowski, C., Elbaum, S., Aftandilian, E., Bowdidge, R.: Programmers'
  {Build} {Errors}: {A} {Case} {Study} (at {Google}).
\newblock In: Proceedings of the 36th {International} {Conference} on
  {Software} {Engineering}, {ICSE} '14, pp. 724--734. ACM, New York, NY, USA
  (2014).
\newblock \doi{10.1145/2568225.2568255}

\bibitem{shu_javapdg:_2013}
Shu, G., Sun, B., Henderson, T., Podgurski, A.: {JavaPDG}: {A} {New} {Platform}
  for {Program} {Dependence} {Analysis}.
\newblock In: Proceedings of the 6th {International} {Conference} on {Software}
  {Testing}, {Verification} and {Validation}, {ICST}'13, pp. 408--415 (2013).
\newblock \doi{10.1109/ICST.2013.57}

\bibitem{shu_mfl:_2013}
Shu, G., Sun, B., Podgurski, A., Cao, F.: {MFL}: {Method}-{Level} {Fault}
  {Localization} with {Causal} {Inference}.
\newblock In: Proceeding of the {Sixth} {International} {Conference} on
  {Software} {Testing}, {Verification} and {Validation}, {ICST}'13, pp.
  124--133 (2013).
\newblock \doi{10.1109/ICST.2013.31}

\bibitem{strug_machine_2012}
Strug, J., Strug, B.: Machine {Learning} {Approach} in {Mutation} {Testing}.
\newblock In: B.~Nielsen, C.~Weise (eds.) Testing {Software} and {Systems}, no.
  7641 in Lecture {Notes} in {Computer} {Science}, pp. 200--214. Springer
  Berlin Heidelberg (2012)

\bibitem{walker_lightweight_2006}
Walker, R.J., Holmes, R., Hedgeland, I., Kapur, P., Smith, A.: A {Lightweight}
  {Approach} to {Technical} {Risk} {Estimation} via {Probabilistic} {Impact}
  {Analysis}.
\newblock In: Proceedings of the {International} {Workshop} on {Mining}
  {Software} {Repositories}, {MSR} '06, pp. 98--104. ACM, New York, NY, USA
  (2006).
\newblock \doi{10.1145/1137983.1138008}

\bibitem{zimmermann_predicting_2008}
Zimmermann, T., Nagappan, N.: Predicting {Defects} {Using} {Network} {Analysis}
  on {Dependency} {Graphs}.
\newblock In: Proceedings of the 30th {International} {Conference} on
  {Software} {Engineering}, {ICSE} '08, pp. 531--540. ACM, New York, NY, USA
  (2008).
\newblock \doi{10.1145/1368088.1368161}

\end{thebibliography}

\end{document}